\documentclass[a4paper,flushrt,twocolumn]{aastex62}
\usepackage{amsmath,amsfonts,amssymb,graphicx,natbib}
\usepackage{indentfirst}    
\usepackage{multirow}
\usepackage[figuresleft]{rotating}
\usepackage{xcolor,colortbl}
\usepackage{epstopdf}
\DeclareGraphicsExtensions{.png,.pdf}
\bibpunct{(}{)}{;}{a}{}{,}

\shorttitle{ISN Gas And PUIs: Large-scale Structures}
\shortauthors{Sok{\'o\l} et al.}
\begin{document}
\title{\textbf{Interstellar Neutral Gas Species and Their Pickup Ions inside the Heliospheric Termination Shock. \\ The Large-scale Structures}}
\correspondingauthor{Justyna M. Sok{\'o\l}}
\email{jsokol@cbk.waw.pl}
\author{Justyna M. Sok{\'o\l}}
\affiliation{Space Research Centre, Polish Academy of Sciences, (CBK PAN), Warsaw, Poland}
\author{Marzena A. Kubiak}
\affiliation{Space Research Centre, Polish Academy of Sciences, (CBK PAN), Warsaw, Poland}
\author{Maciej Bzowski}
\affiliation{Space Research Centre, Polish Academy of Sciences, (CBK PAN), Warsaw, Poland}
%
%
\begin{abstract}
We study the distribution of the interstellar neutral (ISN) gas density and the pick-up ion (PUI) density of hydrogen, helium, neon, and oxygen in the heliosphere for heliocentric distances from inside 1 au up to the solar wind termination shock (TS), both in and out of the ecliptic plane. We discuss similarities and differences in the large-scale structures of the ISN gas and PUIs formed in the heliosphere between various species. We discuss the distribution of ISN gas and PUI densities for two extreme phases of the solar activity cycle, it is the solar minimum and the solar maximum. We identify the location of the ISN gas density cavity of various species. We study the relative abundance ratios of Ne/O, H/He, Ne/He, and O/He for ISN gas and PUIs densities and their variation with location in the heliosphere. We also discuss the modulation of relative abundance ratios of ISN gas and PUIs along the TS. We conclude that the preferable locations for detection of He$^+$ and Ne$^+$~PUIs are in the downwind hemisphere within 1~au, whereas for H$^+$ and O$^+$~PUIs the preferable locations for detection are for distances from Jupiter to Pluto orbits.
\end{abstract}


\section{Introduction \label{sec:intro}}
The neutral component of the local interstellar matter enters the heliosphere freely, which is a cavity around the Sun formed by the interaction of the solar wind with the interstellar matter. Studies of the interstellar neutral (ISN) gas, enabled by measurements from inside the heliosphere, allow us to infer the velocity vector of the heliosphere's motion with respect to the Very Local Interstellar Matter \citep[VLISM; e.g.,][]{bzowski_etal:15a} and bring information about the physical state of the VLISM including temperature \citep[e.g.,][]{bzowski_etal:14a, mccomas_etal:15a} and elemental composition \citep[e.g.,][]{bochsler_etal:12a, park_etal:14a}. Moreover, the ISN gas flow vector can be studied indirectly by the measurement of the so-called pickup ions \citep[PUIs; e.g.,][]{vasyliunas_siscoe:76, mobius_etal:15c}. PUIs are ions produced by ionization of the ISN gas by the solar ionization factors, like charge exchange with solar wind particles, impact ionization by solar wind electrons, and ionization by solar EUV flux (see more in \citet{bzowski_etal:13a, sokol_etal:puiIon}). PUIs are also a source population for energetic neutral atoms (ENAs) and anomalous cosmic rays (ACRs), which both enable the study of the physical processes in the inner and outer heliosheath (see, e.g., a comprehensive review by \citet{zank:15a}). 

We study the ISN gas density of hydrogen, helium, oxygen, and neon and the resulting PUI densities from inside 1~au up to the heliospheric solar wind termination shock (TS) both inside and outside the ecliptic plane for the solar minimum and the solar maximum. We apply the methodology developed by \citet{rucinski_bzowski:95b, bzowski_etal:97, tarnopolski_bzowski:09} for the ISN gas modeling in the heliosphere and \citet{fahr_rucinski:89, rucinski_etal:93a, rucinski_etal:03} for the PUI modeling, with the observation-based ionization rates inside the heliosphere \citep{bzowski_etal:13a, bzowski_etal:13b, sokol_etal:16a, sokol_etal:puiIon}. 

Our aim is to study large-scale structures of the ISN gas and PUI densities inside the heliospheric TS. We discuss similarities and differences in the density distributions for various species. We examine implications for heliospheric studies, like, e.g., the location of the ISN gas cavity, the relative abundance ratios of the ISN gas and PUIs, and the variations along TS. We also identify locations in space that are the most preferable for measurements of PUIs of various species.

The adopted models are briefly presented in Section~\ref{sec:model}. The ISN gas density is discussed in Section~\ref{sec:ISNDensity}. The PUI density is presented in Section~\ref{sec:PUIs}. The implications for the heliospheric studies are discussed in Section~\ref{sec:Implications}. The study is briefly summarized in Section~\ref{sec:Summary}.

\section{Model \label{sec:model}}
\subsection{ISN density \label{sec:modelISN}}
We apply the Warsaw~Test~Particle~Model\footnote{The so-called numerical strain of the WTPM was used (nWTPM).} (WTPM; \citet{sokol_etal:15b}) to calculate the density of ISN gas inside the heliosphere. In the WTPM software, the so-called hot model paradigm of the ISN gas distribution \citep{thomas:78, wu_judge:79a, fahr:78, fahr:79} is used with the Maxwell-Boltzmann distribution function assumed in the source region with the thermal speed,\footnote{$u_T=\sqrt{2kT/m}$, where T -- temperature, k -- Boltzmann constant, and m -- mass} which addresses the differences in mass for each considered species. We use different temperatures, bulk speeds, and flow directions for the primary and secondary populations. All particles are traced back from the locations inside the heliosphere up to the TS with the ionization rates and radiation pressure varying both in space and in time along the trajectories of the atoms.

The inflow parameters (velocity vector and temperature) of the primary ISN gas population are adopted from the results of the analysis of the IBEX-Lo direct-sampling of ISN~He flow after \citet{bzowski_etal:15a}: flow longitude $\lambda_{\mathrm{ISN}}=255.745\degr$ and latitude $\phi_{\mathrm{ISN}}=5.169\degr$ in ecliptic J2000 coordinates, speed $v_\mathrm{B}=25.784$~km~s$^{-1}$, and temperature $T_{\mathrm{B}}=7443$~K. They are identical for ISN H, He, Ne, and O. For secondary He population, we used the results from the study by \citet{kubiak_etal:16a}: $\lambda=251.57\degr$, $\phi=11.95\degr$, $v=11.28$~km~s$^{-1}$, $T=9480$~K, with the abundance with respect to the primary population equal to 0.057. Following \citet{kowalska-leszczynska_etal:18b}, the density, temperature, and speed of the secondary H population were adopted after \citet{bzowski_etal:08a} with the inflow direction identical as for the secondary He: $\lambda=251.57\degr$, $\phi=11.95\degr$, $v=18.744$~km~s$^{-1}$, $T=16300$~K, and with the abundance to the primary population equal to 1.75. In the case of ISN He and H, the local densities are calculated as a sum of the densities of the primary and secondary populations \citep{bzowski_etal:08a} and the total density is presented afterwards. The adopted densities of the species in the source region (upwind at TS) are presented in Table~\ref{tab:normF}.

\subsection{Ionization rates \label{sec:modelIon}}
The most relevant ionization processes for ISN gas inside the heliopshere are charge exchange with solar wind particles, photoionization, and electron impact ionization. Various species are prone to various ionization processes as discussed by \citet{sokol_etal:puiIon}. Thus, the modulation of ISN gas density, as well as the PUI production rates, inside the heliosphere vary between species. In this study, if not stated otherwise, by ionization rates we mean the total ionization rates as a sum of charge exchange with the solar wind particles (protons and alpha particles in the case of He), photoionization, and electron impact ionization. The ionization rates applied in the calculations are derived from the available observations of the solar wind and the solar EUV flux. The evolution of the solar wind proton speed and density with time and heliographic latitude is adopted after \citet{sokol_etal:13a}. The model developed by \citet{sokol_etal:13a} is based on the solar wind in-situ in-ecliptic measurements collected by the OMNI database \citep{king_papitashvili:05} and solar wind speed derived from the interplanetary scintillation observations conducted by the Institute for Space-Earth Environmental Research \citep[ISEE, Nagoya University, Nagoya, Japan;][]{tokumaru_etal:10a, tokumaru_etal:12b}. The photoionization rate is assess based on the solar EUV flux measurements and a system of solar EUV proxies as discussed by \citet{bzowski_etal:13a, bzowski_etal:13b, sokol_etal:puiIon}. The assessment of the electron impact ionization rates follow the methodology developed by \citet{bzowski:08a}. The radiation pressure model used in the calculation of ISN~H is based on the study by \citet{kowalska-leszczynska_etal:18a}.

The total ionization rate model adopted in this study reproduces variations of the ionization factors with time on Carrington rotation time-scale in the ecliptic plane and the solar wind related variations out of the ecliptic plane are modeled with temporal resolution of one year. The solar wind density and the photoionization rates decrease with $r^{-2}$ of the distance to the Sun, and the solar wind speed in kept unchanged with the distance from the Sun.

\subsection{PUIs \label{sec:modelPUIs}}
The in-situ measurements of the ISN gas are challenging due to instrumental limitations, much easier to detect are ions, which are more energetic and have higher fluxes. Thus, the ISN gas flow direction is very often studied by means of the PUIs created in the solar wind. Having computed the ISN density in the heliosphere, we follow the methodology proposed by \citet{fahr_rucinski:89, rucinski_etal:93a, rucinski_etal:03} to calculate the expected PUI production rate, flux, and next the PUI density for the time-dependent ionization rates. However, in our calculations the ionization rates vary in time, distance from the Sun, and heliographic latitude. The total flux of PUIs,  $F_{\mathrm{PUI}}$, at a location $\vec{R}=\left(R,\lambda_{\mathrm{ecl}}, \phi_{\mathrm{ecl}}\right)$, where $R$ is a distance from the Sun, $\lambda_{\mathrm{ecl}}$ is ecliptic longitude, and $\phi_{\mathrm{ecl}}$ is ecliptic latitude, and time $t$, is given by the formula:
\begin{equation}
F_{\mathrm{PUI}}\left(\vec{R}, t\right) = \frac{1}{R^2}\int\limits_{r_0}^{R} n\left(\vec{r},t\right)\beta \left(\vec{r},t\right) r^2 \mathrm{d}r,
\label{eq:puiFlux}
\end{equation}
where $n\left(\vec{r},t\right)$ is the ISN gas density and $\beta \left(\vec{r},t\right)$ is the total ionization rate at time $t$ and at a location $\vec{r}$ defined by distance from the Sun $r$, longitude $\lambda$, and latitude $\phi$ of every point along the line of sight. The product of the ISN density and the ionization rate is called the PUI production rate afterwards. The integration goes from a distance $r_0$ to $R$ along a radial line, where $r_0$ is the solar radius. The density of PUIs at location $\vec{R}$ is calculated by dividing the PUI flux (Equation~\ref{eq:puiFlux}) by the local speed of the solar wind $V_{\mathrm{sw}}\left(\vec{R} \right)$:
\begin{equation}
n_{\mathrm{PUI}}\left(\vec{R},t\right) = F_{\mathrm{PUI}}\left(\vec{R},t\right)/V_{\mathrm{sw}}\left(\vec{R},t\right).
\label{eq:puiDens}
\end{equation}

Estimates of the ionization rates for distances closer than 0.3~au to the Sun, due to the lack of available data, is uncertain. However, we must have an estimate for the ISN gas densities in this region because they significantly contribute to the PUI population within several astronomical units from the Sun. In consequence, we assumed an extrapolation of ionization rates for $r<0.3$~au and we calculate the ISN gas density down to 0.01~au for He and Ne, and 0.1~au for H and O, and next we linearly extrapolate the density to the solar radius. Since the estimates for distances smaller than 0.3~au are based on extrapolations and makes estimates for the PUI density there uncertain, the results are presented afterwards from 0.3~au outward. 

In our model, we assume that PUIs originate solely from ionization of the ISN gas by the solar wind and the solar EUV radiation and we do not include separately the term resulting from the charge exchange between the already existing PUIs and ISN gas as studied by \citet{zank_etal:18a}. \citet{zank_etal:18a} studied the PUI-mediated solar wind flow inside the heliosphere including solar wind, PUIs, interplanetary magnetic field, and low-frequency turbulence to reproduce the PUI observations made by \textit{New Horizons} and \textit{Voyager}~2. Their estimates of H$^{+}$~PUIs agree qualitatively with the estimates from the model discussed in this paper (compare Figure~1 in \citet{zank_etal:18a} with Figure~\ref{fig:PUIDensInflowRadial} here).

The solar wind slows down throughout the heliosphere due to momentum loading due to PUIs injected by charge exchange reaction and the mass loading effect caused by the photoioinization and electron impact ionization \citep{isenberg:86, lee_etal:09a}, as confirmed experimentally by \textit{Voyager}~2 measurements \citet{richardson_etal:08a,richardson_etal:08b}. In consequence, the density of PUIs at TS proportionally increases. But, according to the continuity equation the plasma density decrease is reduced by the slow-down of the solar wind as it results from the charge exchange, as discussed by \citet{bzowski_etal:13a}. Thus, in our calculations the PUI flux is assessed correctly, but the PUI density might be underestimated because of the assumption of the unchanged solar wind speed. However, most of the PUI density is produced at much closer distances to the Sun compared to the TS locations (see, e.g., Figure~\ref{fig:PUIDensInflowRadial} and discussion in \citet{zank_etal:18a}), where the solar wind slow-down effect can be neglected \citep{bzowski_etal:13a}. Additionally, the magnitude of the predicted and measured solar wind slow-down before TS is of the order of the uncertainty of the model constituents we use. Nevertheless, this topic needs further investigation in the future to consistently include the mass and momentum loading effects in the ISN density calculation with the assumption of a hot model paradigm with the ionization rates and radiation pressure variable in time and in space along particle trajectories and PUI modeling.

In our study we also assume an immediate pitch-angle scattering of PUIs and thus we do not discuss the transport of PUIs with the local interplanetary magnetic field and the evolution of PUI distribution function discussed by, e.g., \citet{mobius_etal:15c, quinn_etal:16a}. Those effects affect the determination of the ISN flow direction from PUI observations, which is not within the scope of the present paper.

\subsection{Calculation grid \label{sec:modelGrid}}
\begin{figure*}[t!]
\includegraphics[width=\textwidth]{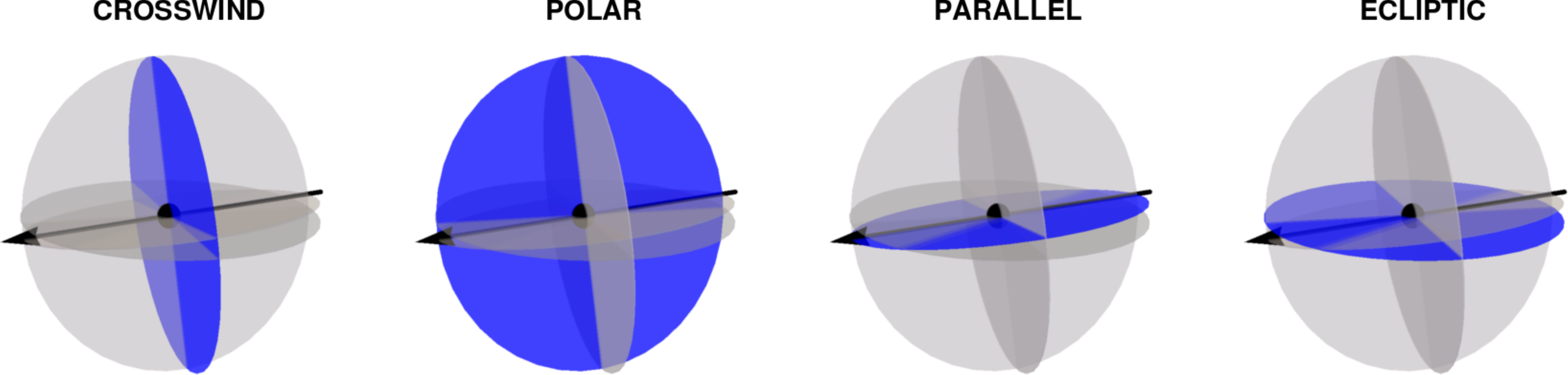}
\centering
\caption{Illustration of the four planes used for the calculations. All four planes are shown in gray at each panel and individual planes are presented in blue. See the text for the definition of the planes. The black dot in the middle of each panel stands for the Sun. The black arrow illustrates the flow direction of the ISN~gas, pointing from upwind to downwind. \label{fig:planes3D}}
\end{figure*}

We carry out the calculations in planes related to (1) the ISN gas flow direction ($\lambda_{\mathrm{ISN}},\phi_{\mathrm{ISN}}$), (2) the variation of the solar ionizing factors, which vary with heliographic latitude, and (3) the geometry of available observations, most of which are limited to the ecliptic plane. We define the following four Sun-centered planes with the ISN flow vector expressed in heliographic coordinates $\left(\vec{V}_{\mathrm{ISN}}^{\mathrm{hel}}\right)$: 
\begin{itemize}
\item{Crosswind: the plane to which the inflow vector is a normal vector $\vec{N}_{\mathrm{crosswind}}=\vec{V}_{\mathrm{ISN}}^{\mathrm{hel}}$.}
\item{Polar: the plane that contains the inflow vector and the north and south heliographic poles; the normal vector of this plane is defined as $\vec{N}_{\mathrm{polar}}=\vec{Z}_{\mathrm{N}}^{\mathrm{hel}} \times \left( -\vec{V}_{\mathrm{ISN}}^{\mathrm{hel}} \right)$, where $\vec{Z}_{\mathrm{N}}^{\mathrm{hel}}=\{0,0,1\}$ is the north heliographic pole.}
\item{Parallel: the plane that contains the inflow vector and the normal vector of the polar plane; the normal vector of this plane is defined as $\vec{N}_{\mathrm{parallel}}=\vec{N}_{\mathrm{polar}} \times \left( -\vec{V}_{\mathrm{ISN}}^{\mathrm{hel}} \right)$.}
\item{Ecliptic: the ecliptic plane, the normal vector of this plane is directed toward the north ecliptic plane $\vec{N}_{\mathrm{ecliptic}}=\vec{Z}_{\mathrm{N}}^{\mathrm{ecl}} = \{0,0,1\}$.}
\end{itemize}
The planes are illustrated in Figure~\ref{fig:planes3D}.

\begin{figure}[t!]
\includegraphics[scale=0.35]{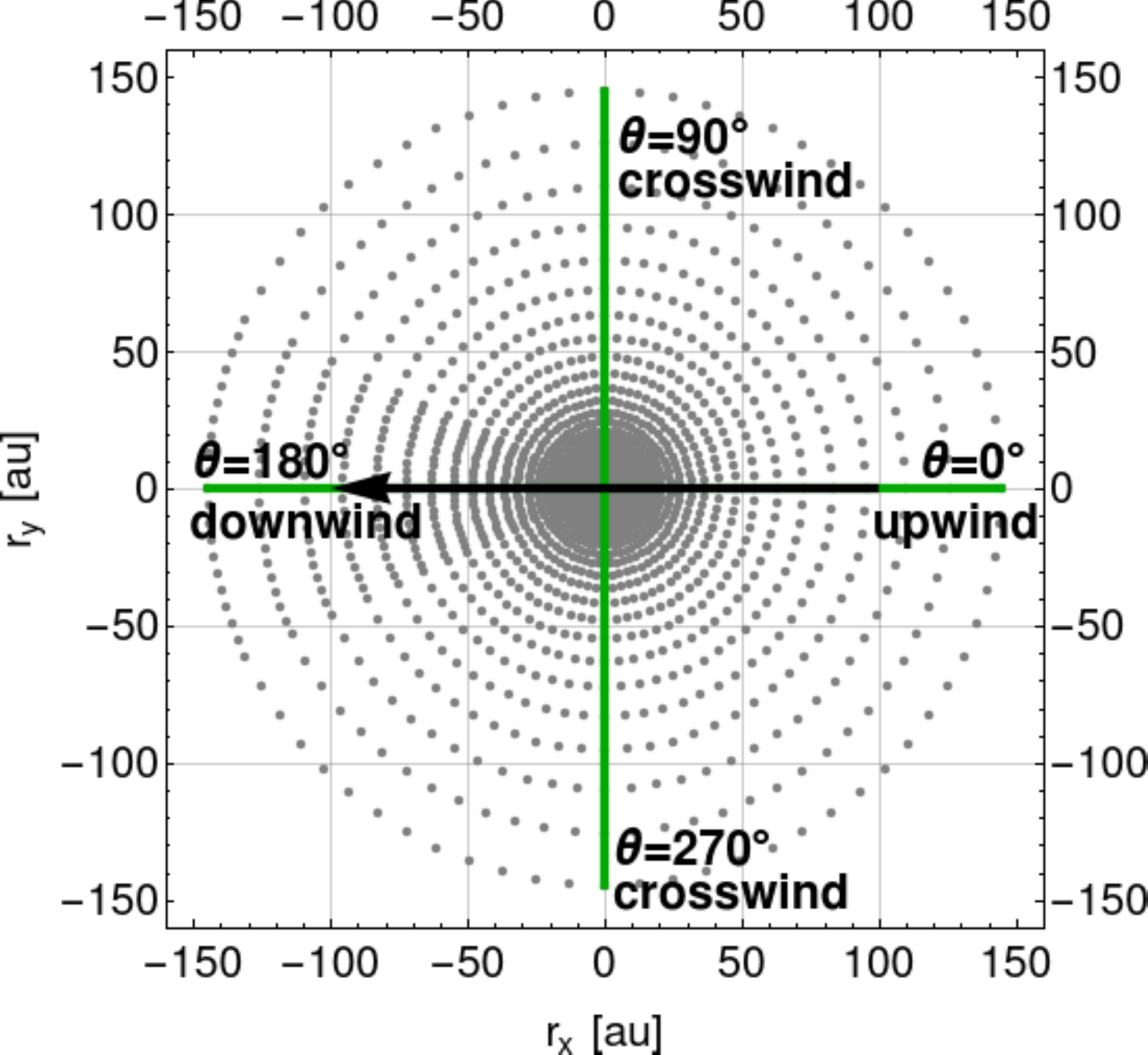}
\centering
\caption{Illustration of the calculation grid for the parallel plane. The arrow indicates the ISN gas flow direction, phase angles $\theta$ for the upwind, downwind, and crosswind directions are marked. The upwind/downwind hemisphere is for positive/negative $r_x$ coordinates of heliocentric distance, respectively. \label{fig:planesGrid}}
\end{figure}

The parallel and ecliptic planes are tilted with respect to each other by $\sim8.92\degr$. Within each plane, the calculation grid is organized in concentric circles around the Sun, as illustrated in Figure~\ref{fig:planesGrid}. The resolution of the grid is $5\degr$ along the circles everywhere except about $\pm25\degr$ around the downwind direction, where the resolution is increased to $2.5\degr$ for the polar, parallel, and ecliptic planes (see Figure~\ref{fig:planesGrid}). The increase of the resolution in the downwind region is to better sample the shape and variations of the ISN gas and PUI densities in the focusing cone. Along the radial lines, the points are distributed logarithmically from $\sim0.1$~au to $\sim150$~au; for the case of He and Ne the inner boundary is set to $\sim0.01$~au. The outer boundary is set to the location of TS determined from the global model of the heliosphere by \citet{heerikhuisen_etal:14a} in its recent version by \citet{zirnstein_etal:16b}. In this model, based on the proton density change, we found the distance to TS equal to $\sim70$~au upwind, $\sim 116$~au downwind, and about 90~au toward the poles.

\begin{figure}[t!]
\begin{tabular}{cc}
 \includegraphics[scale=0.35]{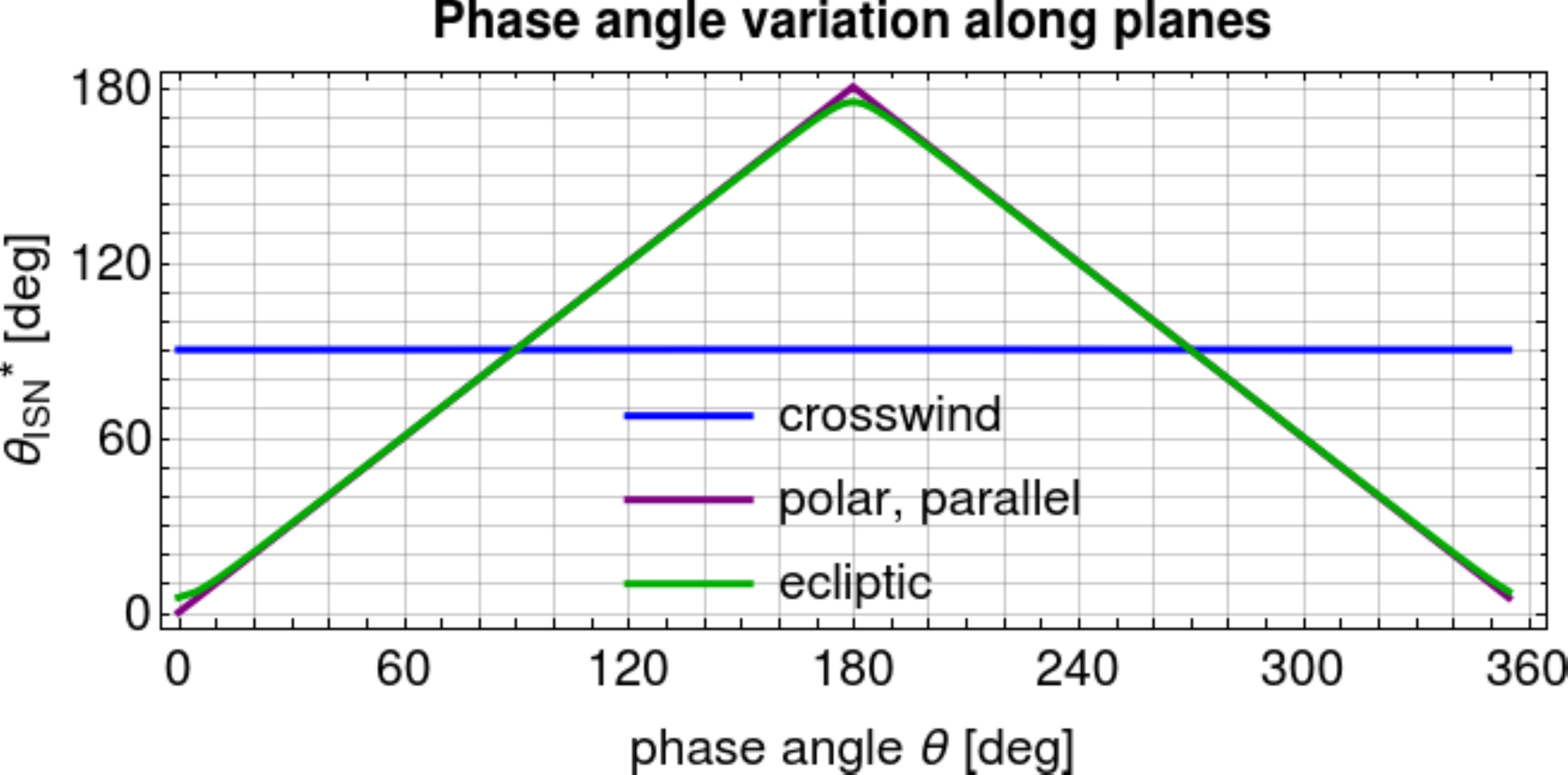}\\
 \includegraphics[scale=0.35]{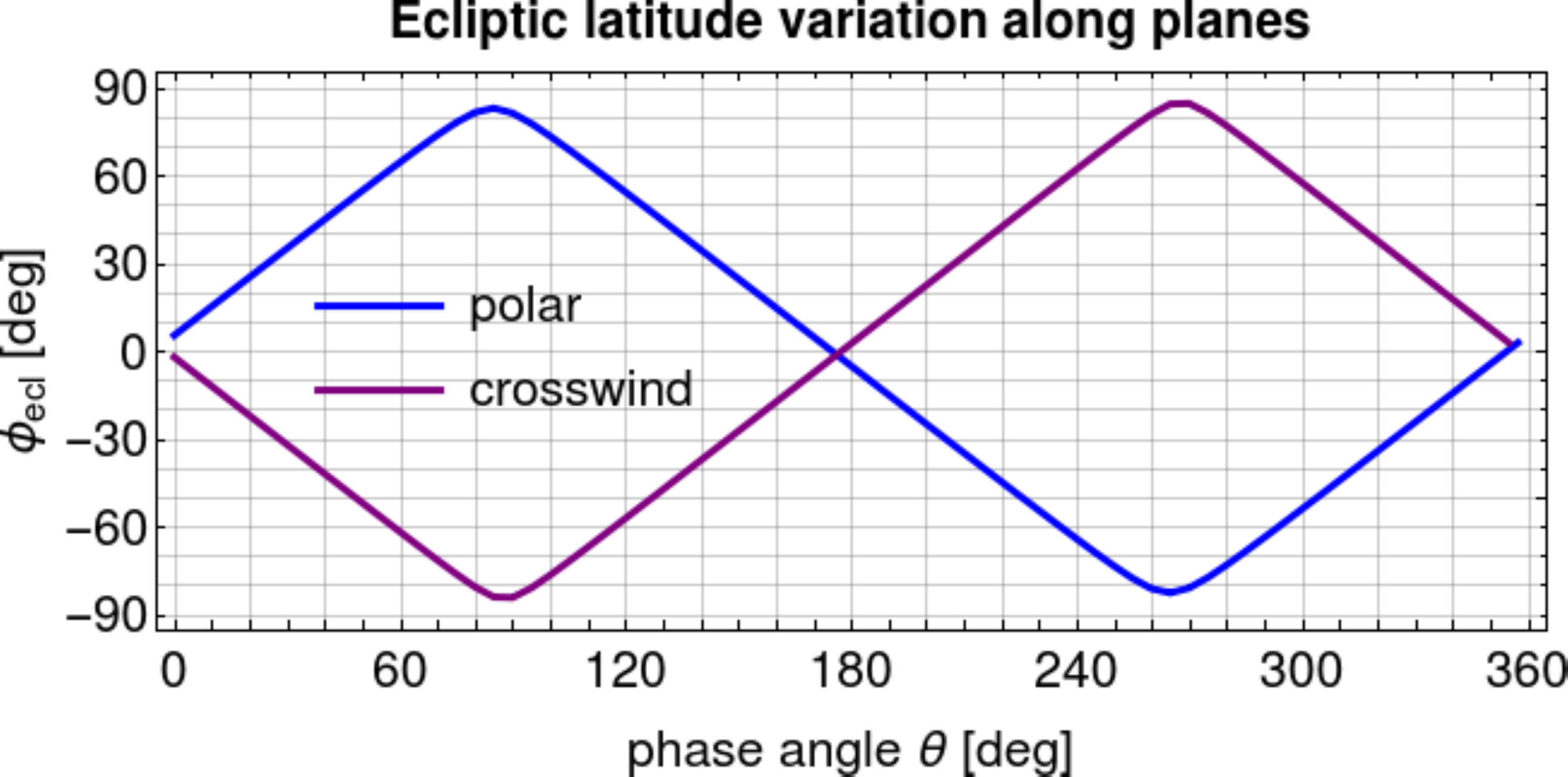}\\
 \includegraphics[scale=0.35]{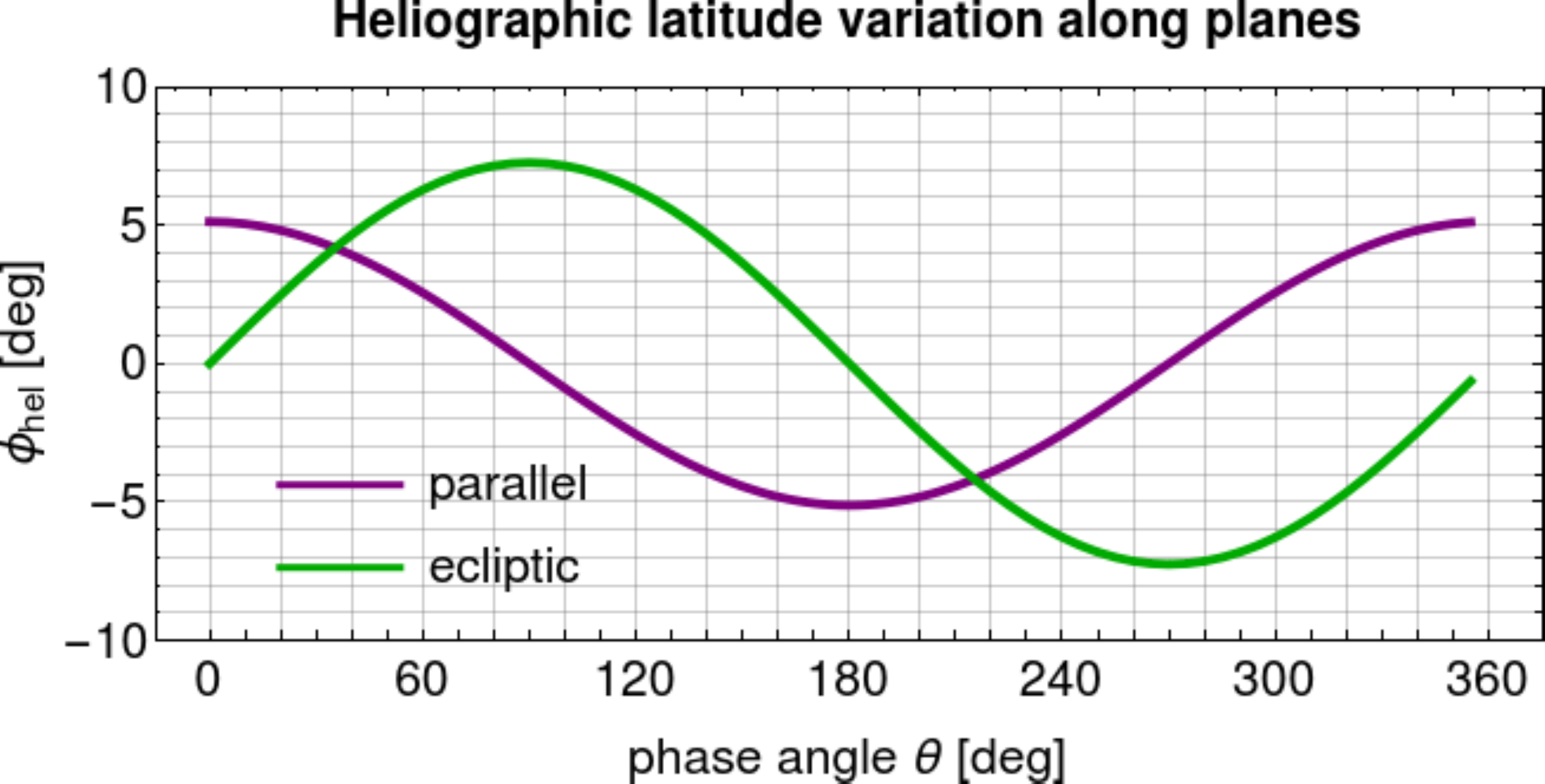}\\
\end{tabular}
\centering
\caption{Top: relation between the phase angle $\theta$ along the planes and the angular distance to the inflow direction $\left( \theta_{\mathrm{ISN}}^* \right)$. Middle: ecliptic latitude $\left(\phi_{\mathrm{ecl}} \right)$ variations as a function of phase angle $\theta$ along the polar and crosswind planes. Bottom: heliographic latitudes $\left(\phi_{\mathrm{hel}} \right)$  of the points in the parallel and ecliptic planes as a function of phase angle $\theta$. \label{fig:planePhAng}}
\end{figure}

The locations of points along each circle are parameterized by the phase angle $\theta \in \left[0\degr,360\degr\right]$ (see Figures~\ref{fig:planesGrid} and \ref{fig:planePhAng}). The phase angle is defined with respect to the ISN gas inflow direction for the polar and parallel planes. The parallel and polar planes contain the flow direction and thus share two points, $\theta=0\degr$ and $\theta=180\degr$. For the ecliptic plane, $\theta=0\degr$ is defined for the direction of $(\lambda_{\mathrm{ecl}},\phi_{\mathrm{ecl}})=(\lambda_{\mathrm{ISN}},0\degr)$. For the crosswind plane, the point $\theta=0\degr$ is counted from the arbitrarily selected point $(\lambda, \phi) = (345.5\degr, -2.3\degr)$ in the ecliptic coordinates. For the polar, parallel, and ecliptic planes, the phase angle increases with the increase of the ecliptic longitude of the points in the grid. Figure~\ref{fig:planePhAng} illustrates the relation between the phase angle $\theta$ and the angle with respect to the inflow direction $\theta_{\mathrm{ISN}}^*$ (top panel), the variation of the ecliptic latitude as a function of phase angle for the polar and crosswind planes (middle panel), and the variation of the heliographic latitude as a function of phase angle along the parallel and ecliptic planes (bottom panel).


\subsection{Moments in time \label{sec:modelTime}}
We study differences between the solar minimum and the solar maximum conditions comparing two moments in time, the solar minimum in 1996 and the solar maximum in 2001. We selected these two years as the most representative for the extreme phases of the solar activity. The results presented are snapshots of the ISN gas and PUI density distributions inside the heliosphere for these two moments in time. 

\section{ISN gas density \label{sec:ISNDensity}} 
\begin{table*}[t]
\centering
\caption{ISN gas and PUI densities for the upwind (Up) and downwind (Dn) directions for selected distances from the Sun ($r$), calculated for 1996 (except the ISN density upwind at TS). \label{tab:normF}}
\begin{tabular}{c|c|c|c|c|c|c}
 \hline
 	\multicolumn{7}{c}{ISN Gas Density $\left[\mathrm{cm}^{-3}\right]$} \\
 	\hline
 	& \multicolumn{2}{c}{r = 1~au} & \multicolumn{2}{c}{r = 5.25~au} & \multicolumn{2}{c}{r = TS} \\
 	\hline
 	& Up & Dn & Up & Dn & \cellcolor{gray!25} Up\footnote{The densities of ISN species at TS upwind adopted after \citet{bzowski_etal:08a} for H, \citet{witte:04} for the primary ISN He, and \citet{kubiak_etal:16a} for the secondary He, and \citet{slavin_frisch:07a,slavin_frisch:08a} for Ne and O. These are the normalization factors for the ISN gas density used in the paper.} & Dn \\
 	\hline
 H & $1.54 \times 10^{-3}$ & $1.90 \times 10^{-4}$ & $3.53 \times 10^{-2}$ & $9.25 \times 10^{-3}$ & \cellcolor{gray!25} $\mathbf{8.52 \times 10^{-2}}$ & $6.96 \times 10^{-2}$ \\
 He & $1.23 \times 10^{-2}$ & $8.68 \times 10^{-2}$ & $1.38 \times 10^{-2}$ & $6.11 \times 10^{-2}$ & \cellcolor{gray!25} $\mathbf{1.50 \times 10^{-2}}$ & $2.23 \times 10^{-2}$ \\ Ne & $2.60 \times 10^{-6}$ & $1.43 \times 10^{-5}$ & $4.55 \times 10^{-6}$ & $2.46 \times 10^{-5}$ & \cellcolor{gray!25} $\mathbf{5.82 \times 10^{-6}}$ & $1.32 \times 10^{-5}$ \\
 O & $3.93 \times 10^{-6}$ & $3.03 \times 10^{-6}$ & $2.70 \times 10^{-5}$ & $4.39 \times 10^{-5}$ & \cellcolor{gray!25} $\mathbf{5.00 \times 10^{-5}}$ & $7.77 \times 10^{-5}$ \\
 \hline \hline
\multicolumn{7}{c}{PUI Density $\left[\mathrm{cm}^{-3}\right]$} \\
 \hline
 & \multicolumn{2}{c}{r = 1~au} & \multicolumn{2}{c}{r = 5.25~au} & \multicolumn{2}{c}{r = TS} \\
 	\hline
 	& Up & Dn & Up & Dn & Up\footnote{Normalization factors for the PUI density used in the paper.} & Dn \\
 	\hline
 H$^{+}$ & $7.58 \times 10^{-5}$ & $8.88 \times 10^{-6}$ & $7.04 \times 10^{-4}$ & $1.52 \times 10^{-4}$ & $2.26 \times 10^{-4}$ & $9.48 \times 10^{-5}$ \\
 He$^{+}$ & $3.11 \times 10^{-4}$ & $2.11 \times 10^{-3}$ & $5.77 \times 10^{-5}$ & $3.31 \times 10^{-4}$ & $4.42 \times 10^{-6}$ & $5.66 \times 10^{-6}$ \\
 Ne$^{+}$ & $1.10 \times 10^{-7}$ & $4.84 \times 10^{-7}$ & $4.39 \times 10^{-8}$ & $2.47 \times 10^{-7}$ & $4.88 \times 10^{-9}$ & $8.62 \times 10^{-9}$ \\
 O$^{+}$ & $2.65 \times 10^{-7}$ & $1.58 \times 10^{-7}$ & $5.64 \times 10^{-7}$ & $7.75 \times 10^{-7}$ & $1.19 \times 10^{-7}$ & $1.16 \times 10^{-7}$ \\
\end{tabular}
\end{table*}

\begin{figure*}
\includegraphics[width=\textwidth]{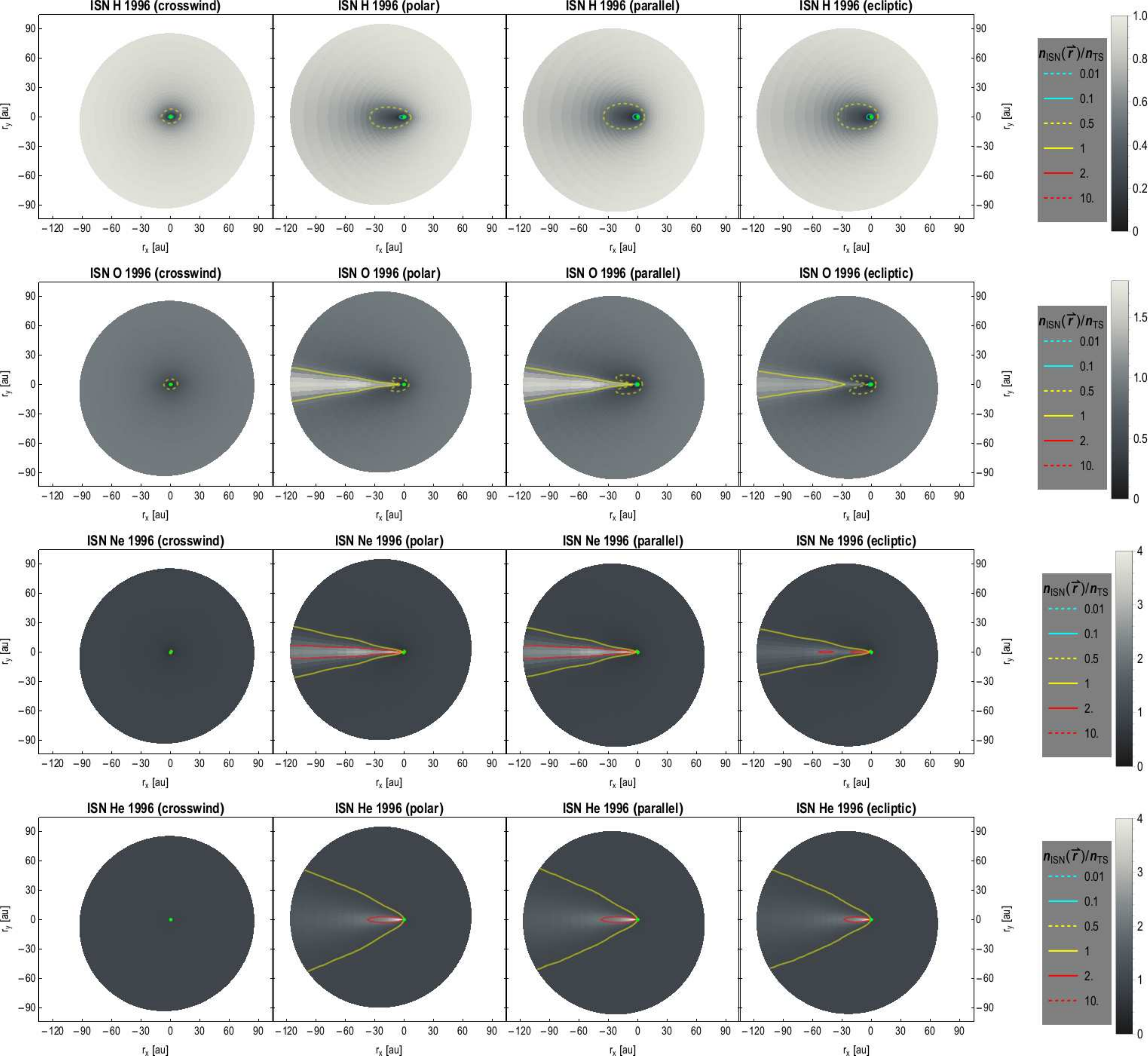}
\centering
\caption{Distribution of the densities of ISN H, O, Ne, and He (from top to bottom) inside TS, normalized to the values upwind at TS (see Table~\ref{tab:normF}), within the crosswind, polar, parallel, and ecliptic planes (from left to right) with the linear scale in the distance from the Sun for 1996. The Sun is at the distance equal to 0, the upwind hemisphere is for positive $r_x$ coordinates, the downwind hemisphere is for negative $r_x$ coordinates. The small green dot in the middle of each panel is a circle placed at 1~au. The intensity is given in the gray scale. Selected isodensity contours are marked at the side bars. (The extended version of the figure uses a logaritymic scale in the distance to the Sun and distances smaller than 0.3~au masked by black disks.) \label{fig:contoursDensMinLin}}
\end{figure*}

\begin{figure*}[t!]
\includegraphics[width=0.7\textwidth]{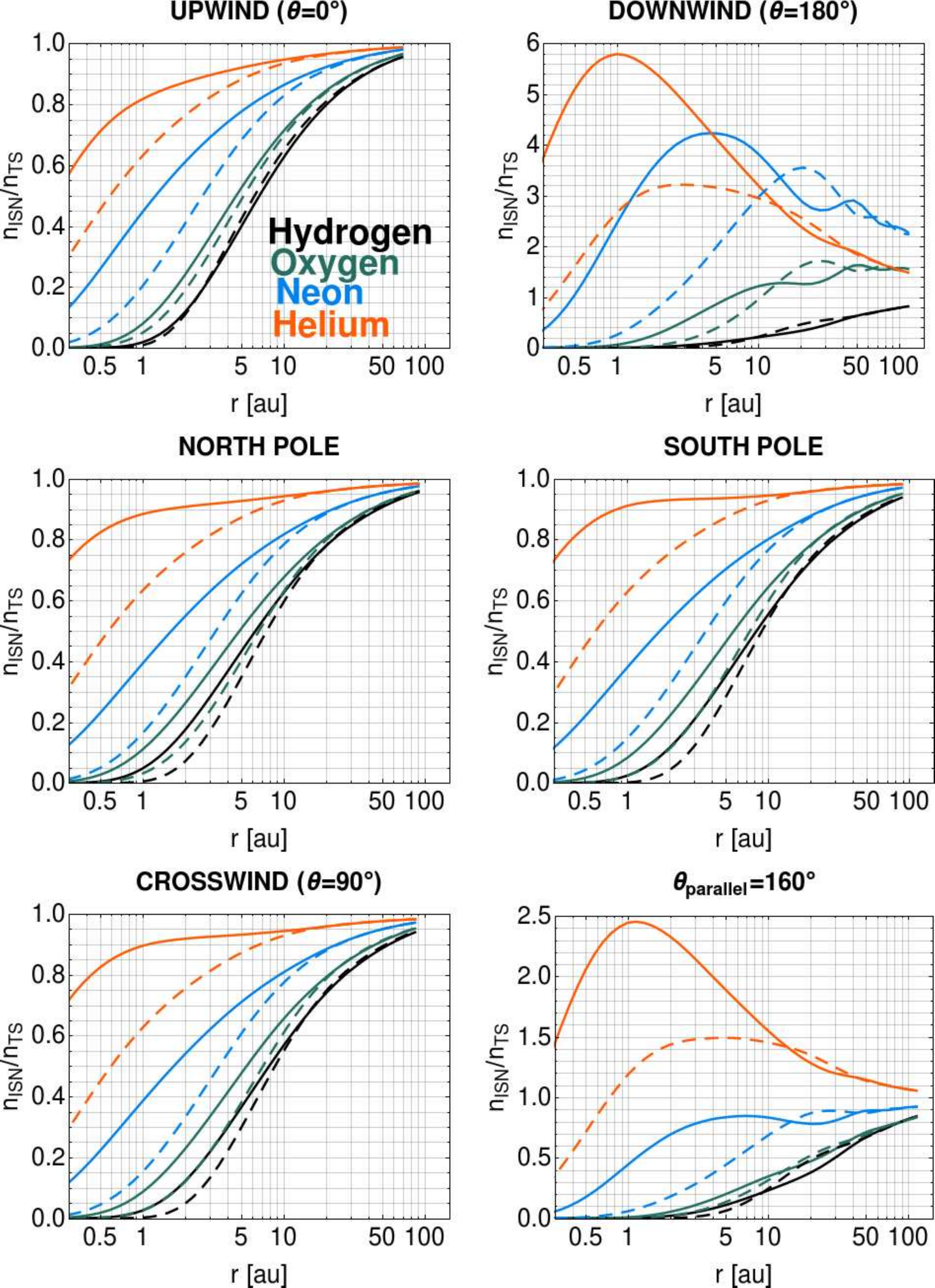}
\centering
\caption{Normalized ISN gas density for H, O, Ne, and He along selected directions in space as a function of distance to the Sun ($r$~[au]). The densities are normalized to the value of the ISN gas density in the upwind direction at TS for each species separately, according to Table~\ref{tab:normF}. The solid lines are for 1996, the dashed lines are for 2001. The direction $\theta=160\degr$ in the parallel plane is an approximate direction for which the ISN Ne and O densities have minimum in the downwind hemisphere (see Figure~\ref{fig:densTheta}). North and south poles are heliographic poles. \label{fig:densInflowRadial}}
\end{figure*}

\begin{figure*}[t!]
\begin{tabular}{cc}
\rotatebox{90}{\textbf{crosswind plane}} & \includegraphics[width=0.95\textwidth]{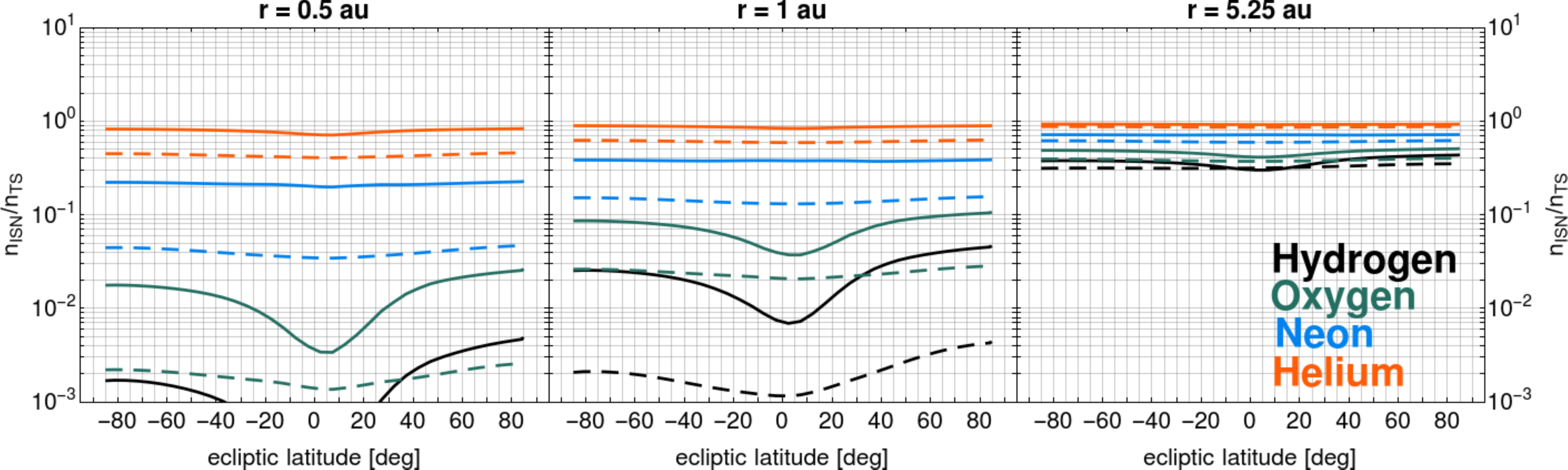} \\
\rotatebox{90}{\textbf{polar plane}} & \includegraphics[width=0.95\textwidth]{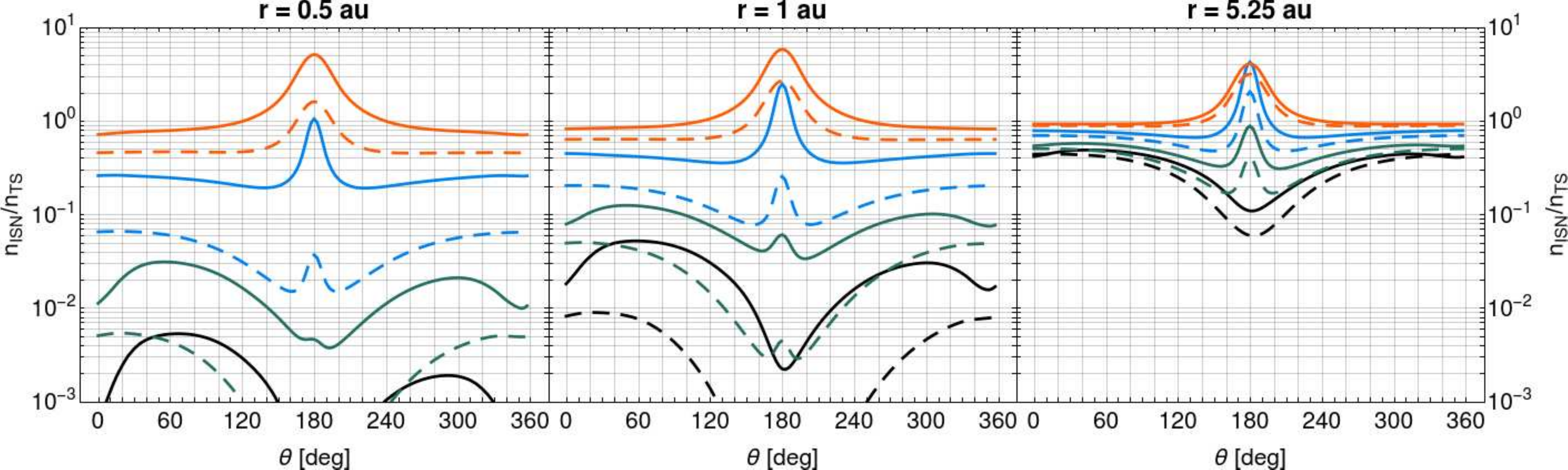} \\
\rotatebox{90}{\textbf{parallel plane}} & \includegraphics[width=0.95\textwidth]{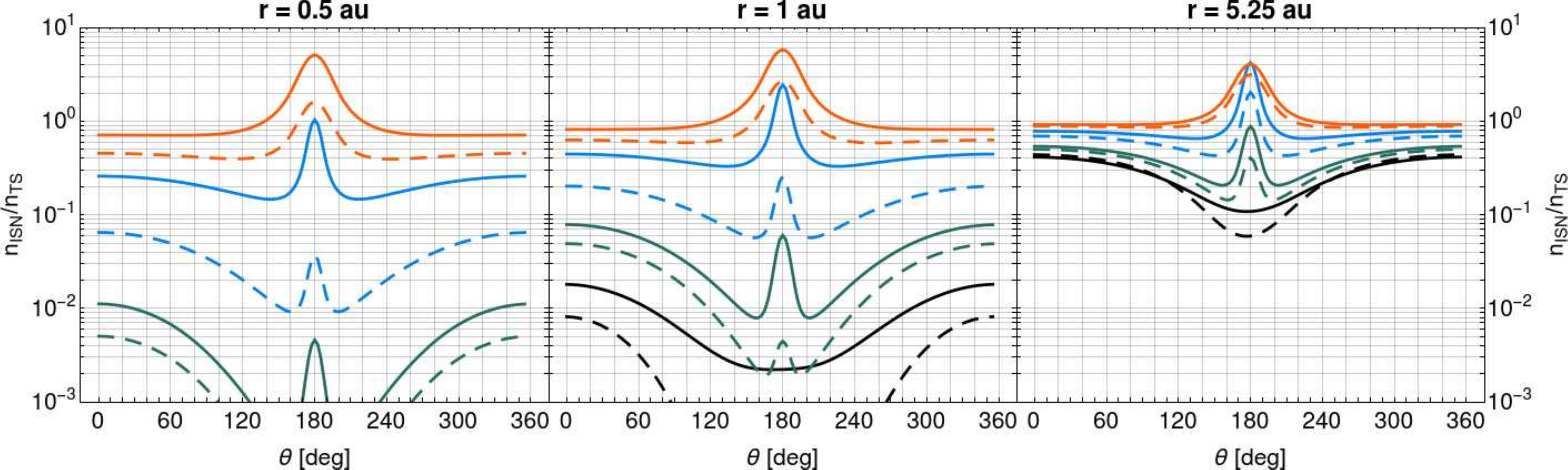} \\
\end{tabular}
\centering
\caption{Normalized ISN density along the crosswind (top), polar (middle), and parallel (bottom) planes for  selected distances from the Sun. The densities are normalized to the value of the ISN gas density in the upwind direction at TS for each species separately, according to Table~\ref{tab:normF}. Solid lines are for 1996, and dashed lines for 2001. Please note that for $r=0.5$~au the ISN~H normalized densities are below the lower limit of the figure and thus not present in the figure. See the text for an explanation. \label{fig:densTheta}}
\end{figure*}

The calculated global distribution of the ISN gas in 1996 is presented in Figure~\ref{fig:contoursDensMinLin}. Each row corresponds to a different species (H, O, Ne, and He) and each column to a different plane (crosswind, polar, parallel, and ecliptic). The density presented is normalized to the ISN density assumed in the upwind direction at TS, according to Table~\ref{tab:normF}. 

The structure of the ISN~H density distribution is quite different from the structure characteristic for the remaining species, as found in the early studies by \citet{axford:72, thomas:78, fahr:78, fahr:79}. Due to the joint action of the ionization processes and the radiation pressure, the ISN~H gas density is significantly reduced at a few astronomical units to the Sun and, in consequence, an ISN~H cavity is created relatively far from the Sun \citep[e.g.,][]{rucinski_bzowski:95b, mccomas_etal:04a}. The ISN~H cavity, as well as cavities for the remaining species, are discussed in more detail in Section~\ref{sec:Cavity}. There is not a maximum of the gas density at the downwind side for ISN~H; instead, the density asymptotically increases outward from the Sun toward the density in the VLISM, as presented in Figure~\ref{fig:densInflowRadial}. 

The density of ISN~H is reduced throughout the heliosphere, whereas for ISN~O, Ne, and He the density is enhanced in a region around the downwind axis known as the cone. As illustrated in Figures~\ref{fig:contoursDensMinLin} and \ref{fig:densInflowRadial}, the cone extends downwind from the Sun up to the TS and farther out. While Figure~\ref{fig:contoursDensMinLin} presents the similarities and differences of the large-scale structures of the ISN gas, the online-only Figure~\ref{fig:contoursDensMinLogr} presents the same quantities, but with distance from the Sun in a logarithmic scale, highlighting the structures close to the Sun. 

A clear, one-peak cone is present for ISN~He density, with the maximum close to 1~au, which makes it easily detectable by instruments in the Earth's orbit regardless of the phase of solar activity. The peak value in 1996 is about twice the value in 2001. Additionally, as Figure~\ref{fig:densInflowRadial} presents, the decay of the density behind the peak is slower/shallower in 1996 than in 2001.

The intensities of ionization losses for H and O are very similar \citep{sokol_etal:puiIon}; however, the resulting distributions of ISN gas are significantly different (Figures~\ref{fig:contoursDensMinLin}, \ref{fig:densInflowRadial}, \ref{fig:densTheta}). This is a result of different thermal speeds for H and O and the action of radiation pressure on H atoms \citep{rucinski:85a}. The main difference is in the downwind hemisphere, where the density cone forms clearly for oxygen and a depletion region forms for hydrogen. As already discussed by \citet{rucinski_bzowski:95b, bzowski_etal:97} and experimentally studied by \citet{mccomas_etal:04a}, the ISN~H is significantly depleted at close distances to the Sun, especially during the solar maximum, when the ionization rates are greater and the radiation pressure exceeds the gravitational force. For the ISN~O density, both the cavity upwind and the density cone downwind are present. 

The structures of the ISN~Ne and He density are similar to each other. Both are focused downwind in the cone region; however, the helium cone is approximately two-fold wider than the neon and oxygen cones (full width at half of the flux maximum in the cone is $\sim 35\degr$ at 1~au for He and $\sim 15\degr$ for Ne and O), see, e.g., Figure~\ref{fig:densTheta}. This is a consequence of a larger value of the ratio of bulk to thermal speeds (the Mach numbers for O and Ne are larger than for He\footnote{The Mach numbers of the primary populations of the ISN species ($M=v_B/\sqrt{\frac{5}{6}\frac{2kT}{m}}$): $M_{\mathrm{H}}=2.55$, $M_{\mathrm{He}}=5.08$, $M_{\mathrm{Ne}}=11.35$, and $M_{\mathrm{O}}=10.15$}). The density enhancement downwind is the greatest in 1996 and for He ($\sim 6$ times wrt upwind value), next for Ne ($\sim 4$) and O ($\sim 1.5$), as shown in Figure~\ref{fig:densInflowRadial}. Additionally, the location of the maximum of the ISN gas density is the closest to the Sun for He (at 1~au), further from the Sun for Ne (at $4 - 6$~au), and next for O (at distances much greater than 10~au, at about $\sim 55$~au). A comparison of the solid and dashed lines in Figure~\ref{fig:densInflowRadial} shows that the distance of the ISN gas density peak in the cone depends on the phase of the solar activity and the peak is further away from the Sun in 2001 than in 1996 for He (at $\sim 3$~au)  and for Ne ($\sim 20$~au). An exception is the peak for ISN~O gas density, for which the maximum is located closer to the Sun in 2001 ($\sim 28$~au) than in 1996  ($\sim 55$~au). The variation in density peak location in time is a consequence of stronger ionization during the solar maximum, which causes that more atoms to be ionized closer to the Sun, and only those that survive are focused downwind. 

The feature of a wavy structure outward from the Sun in the radial variations in the downwind direction (the most pronounced for the ISN~Ne and O density as presented in the upper right panel in Figure~\ref{fig:densInflowRadial}) is the propagation of the ionization wave already reported for ISN~H by \citet{rucinski_bzowski:95b} (see their Figure~4). It is a result of the solar cycle variations of the ionization rates (and the radiation pressure in the case of ISN~H) and the history of these variations accounted for in the calculation of the ISN density. Because the ISN atoms travel through the heliosphere from tens to a few tens of years, they are exposed to periodic solar cycle variations of the ionization strength on their route to the Sun (see Figure~3 in \citet{bzowski_etal:13b}), which results in quasi-periodic modulation of the ISN gas density.

Table~\ref{tab:normF} summarizes the absolute ISN gas densities for the upwind and downwind directions for selected distances from the Sun, calculated for 1996. Additionally, Figure~\ref{fig:densTheta} illustrates the differences between the ISN gas density along the crosswind, polar, and parallel planes between 1996 and 2001 for selected distances from the Sun. In the case of H and O, the ISN gas density variations in the crosswind plane reflect the ionization rate variations with latitude, with smaller densities around the solar equator, where the ionization rates are greater, and the greater densities at higher latitudes, where the ionization rates are smaller (see also Figure~5 in \citet{sokol_etal:puiIon}). The bimodal structure in the crosswind plane (clearly visible at $r=1$~au) becomes uniform with an increase of the heliocentric distance due to decrease of the ionization rates with increasing distance to the Sun. Please note that due to the strong depletion of ISN~H at close distance to the Sun, the ISN~H is almost entirely below the limit of the figure at $r=0.5$~au. The difference in variations of ISN~O density along polar and parallel planes, which share directions $\theta=0\degr$ and $\theta=180\degr$, is a result of different ionization rates along atoms' trajectories within the planes. The more intense variation in the polar plane reflects the latitudinal anisotropy of the ionization rates, which feature a strong modulation at the poles during the solar cycle (see Figure~5 in \citet{sokol_etal:puiIon}).

\section{PUI density \label{sec:PUIs}} 
\begin{figure*}
\includegraphics[width=\textwidth]{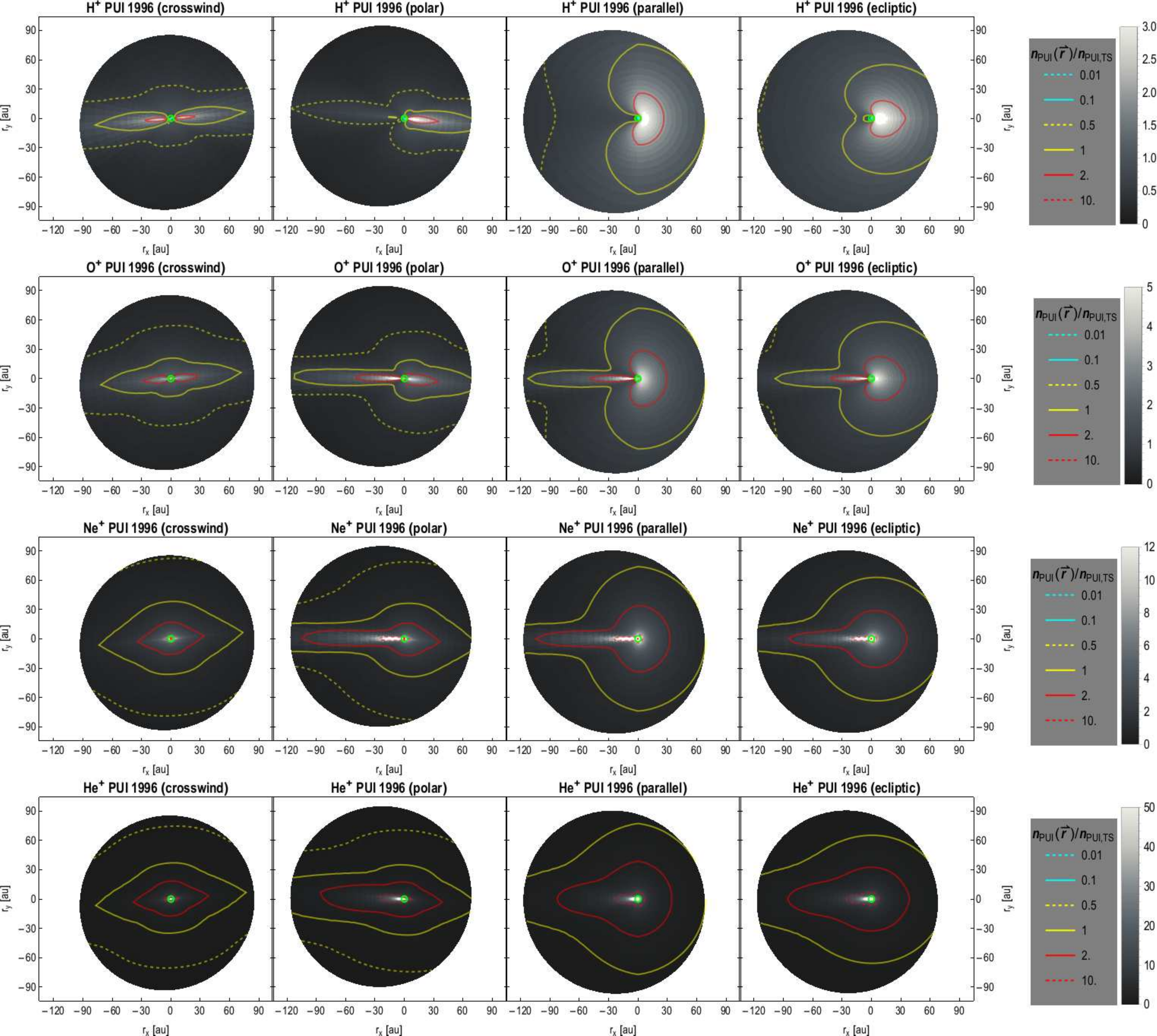}
\centering
\caption{Distribution of the density of H$^+$, O$^+$, Ne$^+$, and He$^+$ PUIs (from top to bottom) inside TS for the crosswind, polar, parallel, and ecliptic planes in 1996. The PUI density is normalized to the PUI density upwind at TS in 1996 for each species separately (see Table~\ref{tab:normF}). The Sun is at the distance equal to 0, the upwind hemisphere is for positive $r_x$ coordinates, the downwind hemisphere is for negative $r_x$ coordinates. The small green dot in the middle of each panel is a circle placed at 1~au. The intensity is given in the gray scale. Selected isodensity contours are marked in the side bars. \label{fig:contoursPUIDensMinLin}}
\end{figure*}

\begin{figure*}
\includegraphics[width=\textwidth]{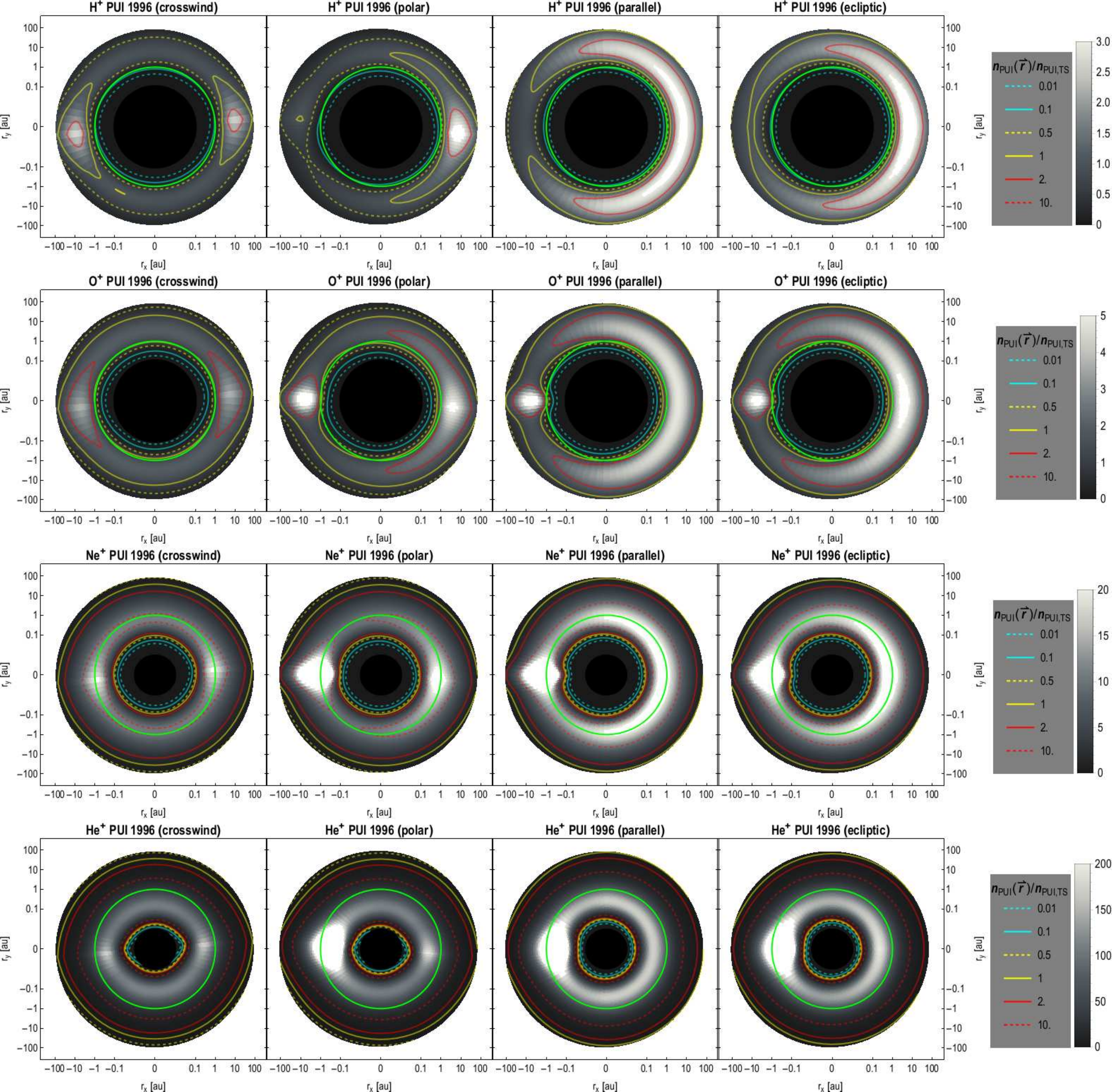}
\centering
\caption{Same as Figure~\ref{fig:contoursPUIDensMinLin} but with logarithmic scale in the distance from the Sun. \label{fig:contoursPUIDensMinLogr}}
\end{figure*}

\begin{figure*}[t!]
\includegraphics[width=0.7\textwidth]{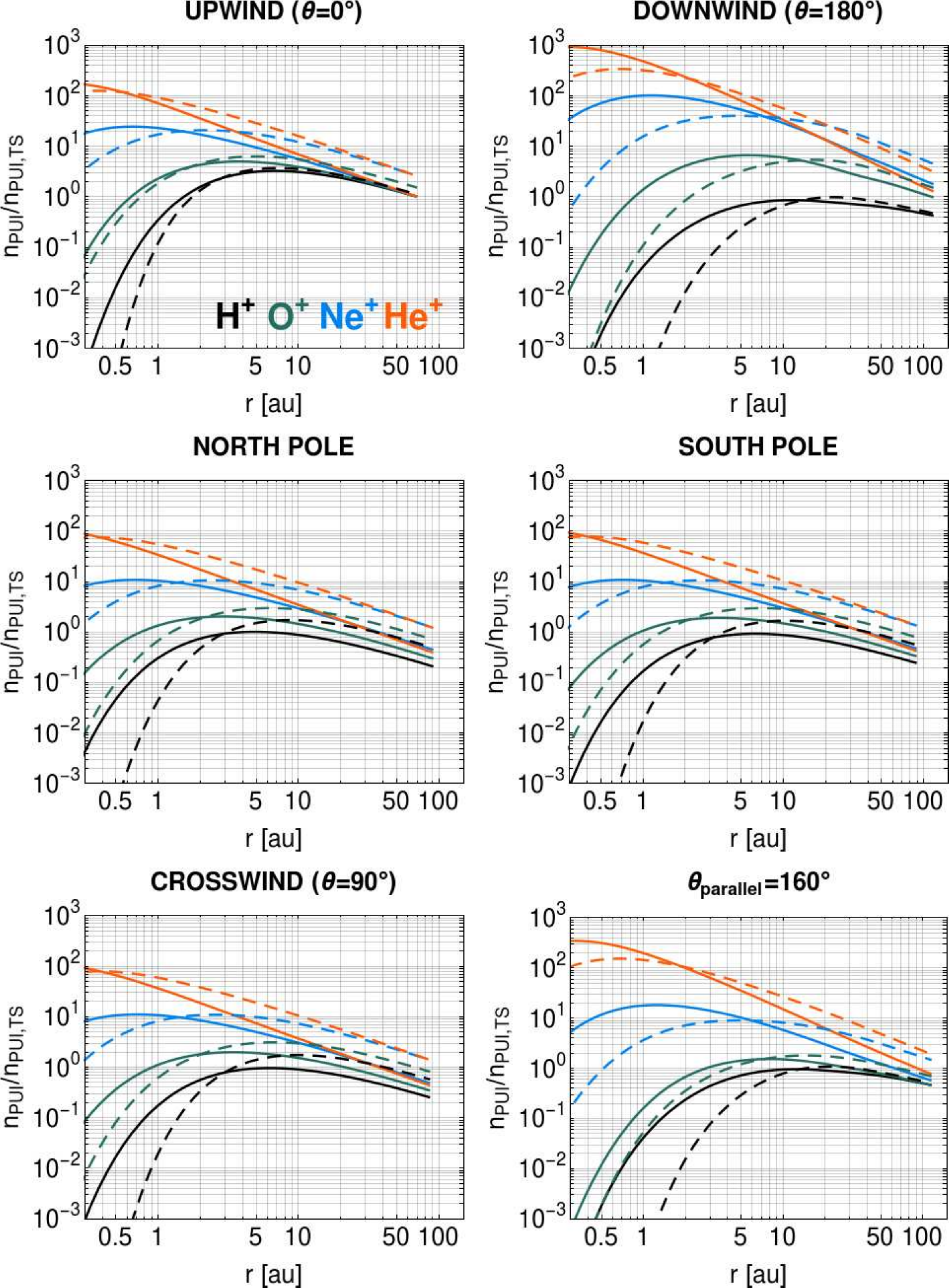}
\centering
\caption{Normalized PUI density for H$^{+}$, O$^{+}$, Ne$^{+}$, and He$^{+}$ along selected directions in space. The normalization is to the PUI density upwind at TS in 1996 for each species separately (see Table~\ref{tab:normF}). The solid lines are for 1996, the dashed lines are for 2001. The direction $\theta=160\degr$ in the parallel plane is an approximate direction for which the ISN Ne and O densities have minimum in the downwind hemisphere (see Figure~\ref{fig:densTheta}). North and south poles are heliographic poles. \label{fig:PUIDensInflowRadial}}
\end{figure*}

\begin{figure*}[t!]
\begin{tabular}{cc}
\rotatebox{90}{\textbf{crosswind plane}} & \includegraphics[width=0.95\textwidth]{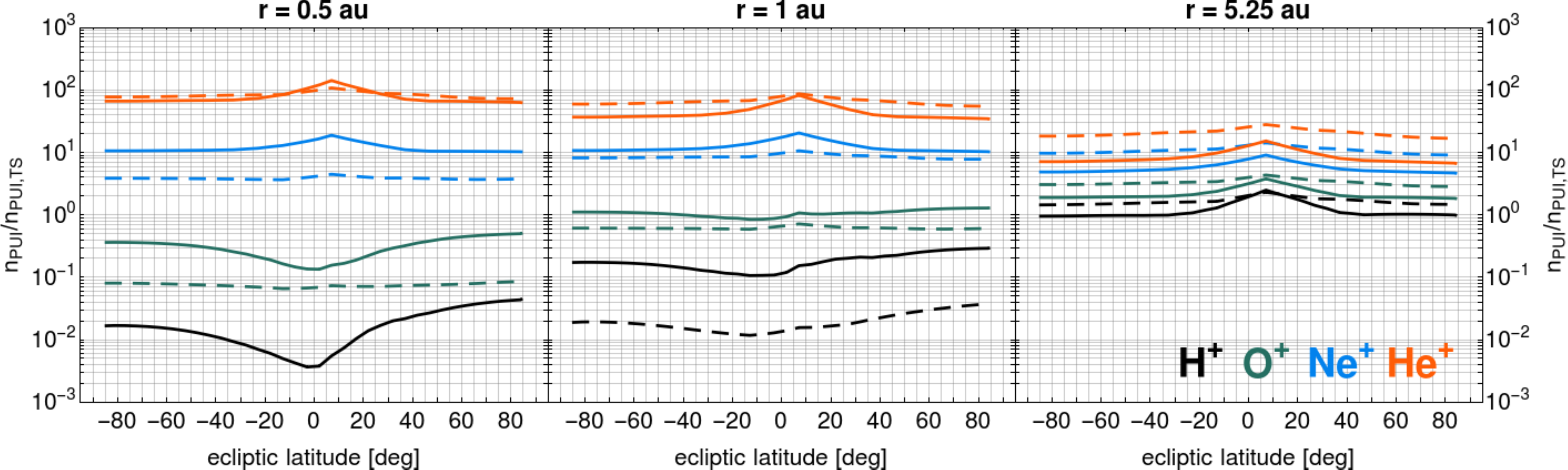} \\
\rotatebox{90}{\textbf{polar plane}} & \includegraphics[width=0.95\textwidth]{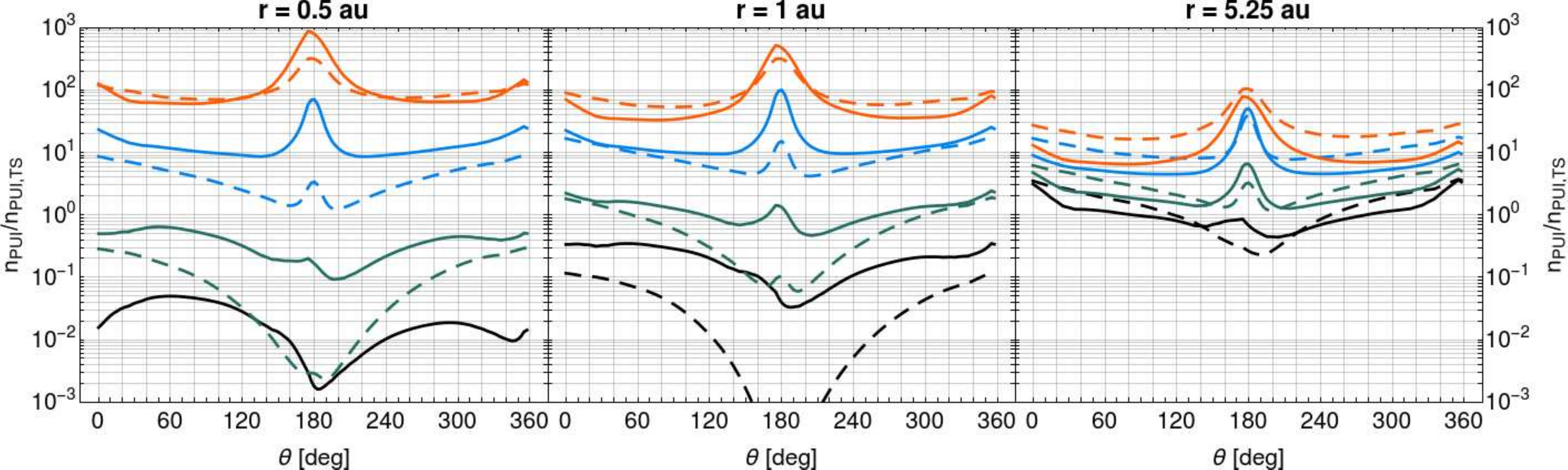} \\
\rotatebox{90}{\textbf{parallel plane}} & \includegraphics[width=0.95\textwidth]{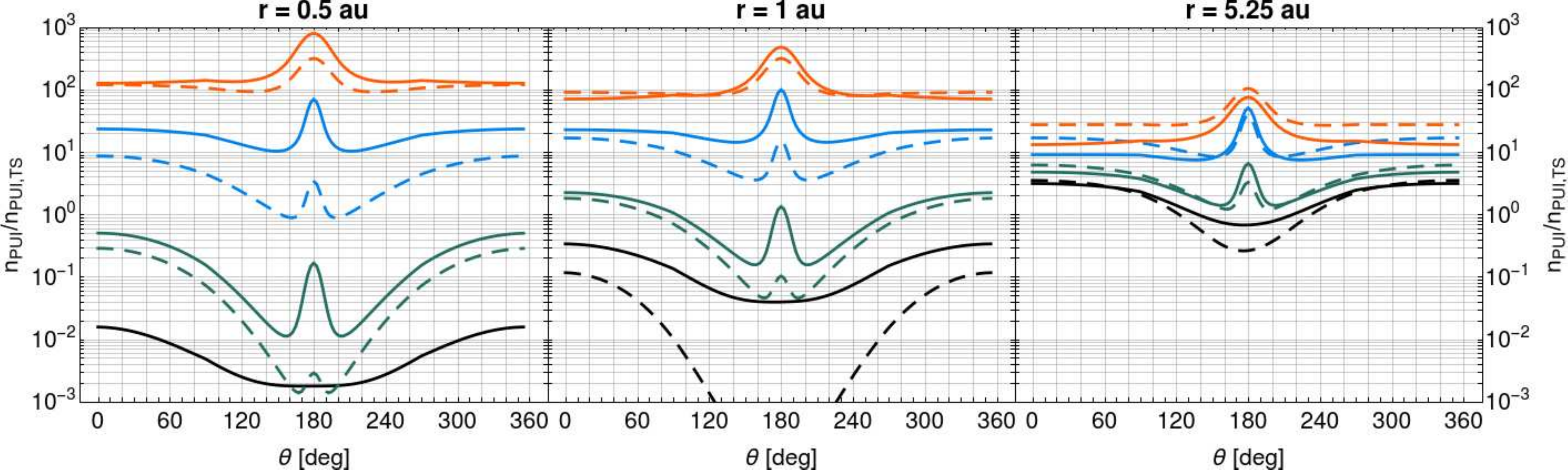} \\
\end{tabular}
\centering
\caption{Normalized PUI density along crosswind (top), polar (middle), and parallel (bottom) planes for selected distances from the Sun. The normalization is to the PUI density upwind at TS in 1996 for each species separately (see Table~\ref{tab:normF}). The solid lines are for 1996, the dashed lines are for 2001. Please note that the black dashed line for H$^{+}$~PUI density at $r=0.5$~au in 2001 is below the lower limit of the figure. \label{fig:densPUITheta}}
\end{figure*}

The PUI densities calculated according to Equation~\ref{eq:puiDens} for H$^+$, O$^+$, Ne$^+$, and He$^+$ are presented in Figure~\ref{fig:contoursPUIDensMinLin} from 0.3~au up to the TS. The PUI distribution is large scale, as is visible in linear scale, but the logarithmic scale for distances from the Sun reveals, additionally, that the structures which are close to the Sun and highlights the differences in PUI density distributions between various species, as illustrated in Figure~\ref{fig:contoursPUIDensMinLogr}. Comparison of Figures~\ref{fig:contoursPUIDensMinLin} and \ref{fig:contoursPUIDensMinLogr} for PUI density distributions with Figures~\ref{fig:contoursDensMinLin} and \ref{fig:contoursDensMinLogr} for the ISN gas density distributions reveals the differences between PUIs and ISN gas from which the PUIs originate. The more variable structures in the PUI density distributions is a consequence of modulation of the PUI production rates by ionization rates. Additionally, the ISN gas density and the ionization rates vary differently as a function of distance from the Sun and latitude \citep{sokol_etal:puiIon}.

In the crosswind plane, the bimodal structure of the PUI density with enhancement around the solar equator is pronounced (Figure~\ref{fig:contoursPUIDensMinLin}). This is a consequence of higher PUI production rates around the solar equator due to the higher ionization rates in this region. In the polar, parallel, and ecliptic planes, the PUI density forms  a cone downwind, similarly as for the ISN density. The PUI cone, however, is formed closer to the Sun than the ISN gas density cone.

Additionally, an enhancement in the PUI density is formed in the upwind hemisphere, known as the crescent (see Figures~\ref{fig:contoursPUIDensMinLin} and \ref{fig:contoursPUIDensMinLogr}). The crescent formation depends on the ionization rates: the higher the ionization, the more pronounced the crescent is. For example, the crescent is present for O$^+$ and H$^+$~PUIs regardless of the phase of the solar activity (see, e.g., Figure~\ref{fig:densPUITheta}). For Ne$^+$~PUIs density the crescent is more pronounced when compared to cone during the solar maximum, when the ionization rates for Ne are greater. For He$^+$~PUIs, the crescent is not as pronounced as for the other species and it is formed inside the Earth orbit. Both the crescent and the cone of PUIs of various species were studied experimentally by, e.g., \textit{STEREO}/Plastic instruments \citep{drews_etal:12a}. 

As presented in Figure~\ref{fig:contoursPUIDensMinLogr} qualitatively and Figure~\ref{fig:PUIDensInflowRadial} quantitatively, for He$^+$ and Ne$^+$ PUIs the enhancement of PUI density forms closer to the Sun than for ISN gas (compared with Figure~\ref{fig:densInflowRadial}). The PUI density peaks downwind are at $\sim 0.2\,(\sim 0.7)$~au for He$^+$, $\sim 1\,(\sim 4.6)$~au for Ne$^+$, $\sim 6\,(\sim 16)$~au for O$^+$, and $\sim 12\,(\sim 24)$~au for H$^+$ in 1996 (2001). Upwind, the PUI density peaks are expected at $<0.3\,(\sim 0.4)$~au for He$^+$, $\sim 0.7\,(\sim 2.3)$~au for Ne$^+$, $\sim 4\,(\sim 5.2)$~au for O$^+$, and $\sim 7$ $(\sim 7)$~au for H$^+$ in 1996 (2001). In the case of H$^+$~PUIs, the highest density is expected upwind (more than 3 times greater than at the TS upwind). For O$^+$~PUIs, the density enhancement is similar for both the crescent and the cone (from 5 to 6 times greater than upwind at the TS), while the crescent peak is closer to the Sun than the cone peak. For Ne$^+$~PUIs, the density enhancement is the greatest downwind in 1996 (about 100 times wrt to upwind at the TS) and reduces to about 40 times in 2001 (wrt to the upwind value at the TS in 1996). For the Ne$^+$~PUI crescent, the enhancement is of a factor of $\sim20$ for both 1996 and 2001. In the case of He$^+$~PUIs, the density enhancement is the greatest downwind, being of a factor of almost 10$^3$ in 1996 and $3\times10^2$ in 2001. Table~\ref{tab:normF} summarizes the absolute densities of PUIs for the upwind and downwind directions for selected distances from the Sun.

Additionally, a comparison of solid and dashed lines in Figures~\ref{fig:PUIDensInflowRadial} and \ref{fig:densPUITheta} illustrates the variation of the PUI density with the phase of the solar activity. The expected PUI density is smaller closer to the Sun in 2001 and higher at far distances from the Sun than in 1996, see the heliocentric distance of the intersection of the solid and dashed lines in Figure~\ref{fig:PUIDensInflowRadial} and interlacing of these lines in Figure~\ref{fig:densPUITheta} as a function of phase angle along the planes. The stronger ionization rates in 2001 (during the solar maximum) cause a smaller ISN gas density at closer distances to the Sun (Figure~\ref{fig:densInflowRadial} and \ref{fig:densTheta}). In consequence, fewer ions are produced at close distances to the Sun and more ISN atoms are ionized at further distances from the Sun. As a result, the maximum of PUI density moves outward from the Sun. Additionally, the variation of the  PUI density both as a function of latitude and as a function of angular distance from the upwind direction, holds throughout the heliosphere with only the magnitude modulated between 1996 and 2001 as illustrated in Figure~\ref{fig:densPUITheta} (see further discussion in Section~\ref{sec:varTS}).

\section{Discussion and conclusions \label{sec:Implications}} 

\subsection{Cavity\label{sec:Cavity}}
\begin{figure*}[t!]
\begin{tabular}{cc}
\includegraphics[width=0.3\textwidth]{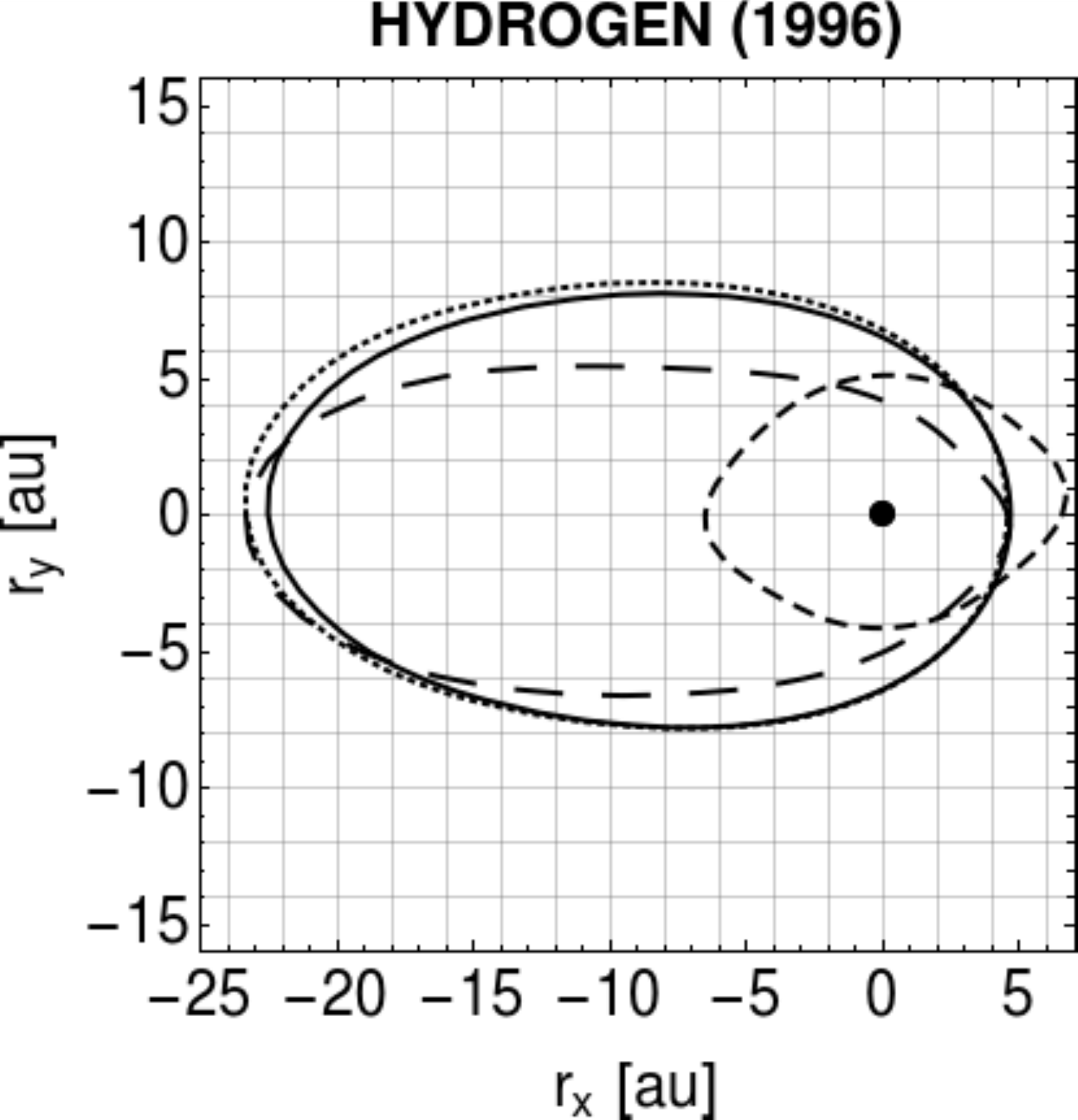} & \includegraphics[width=0.49\textwidth]{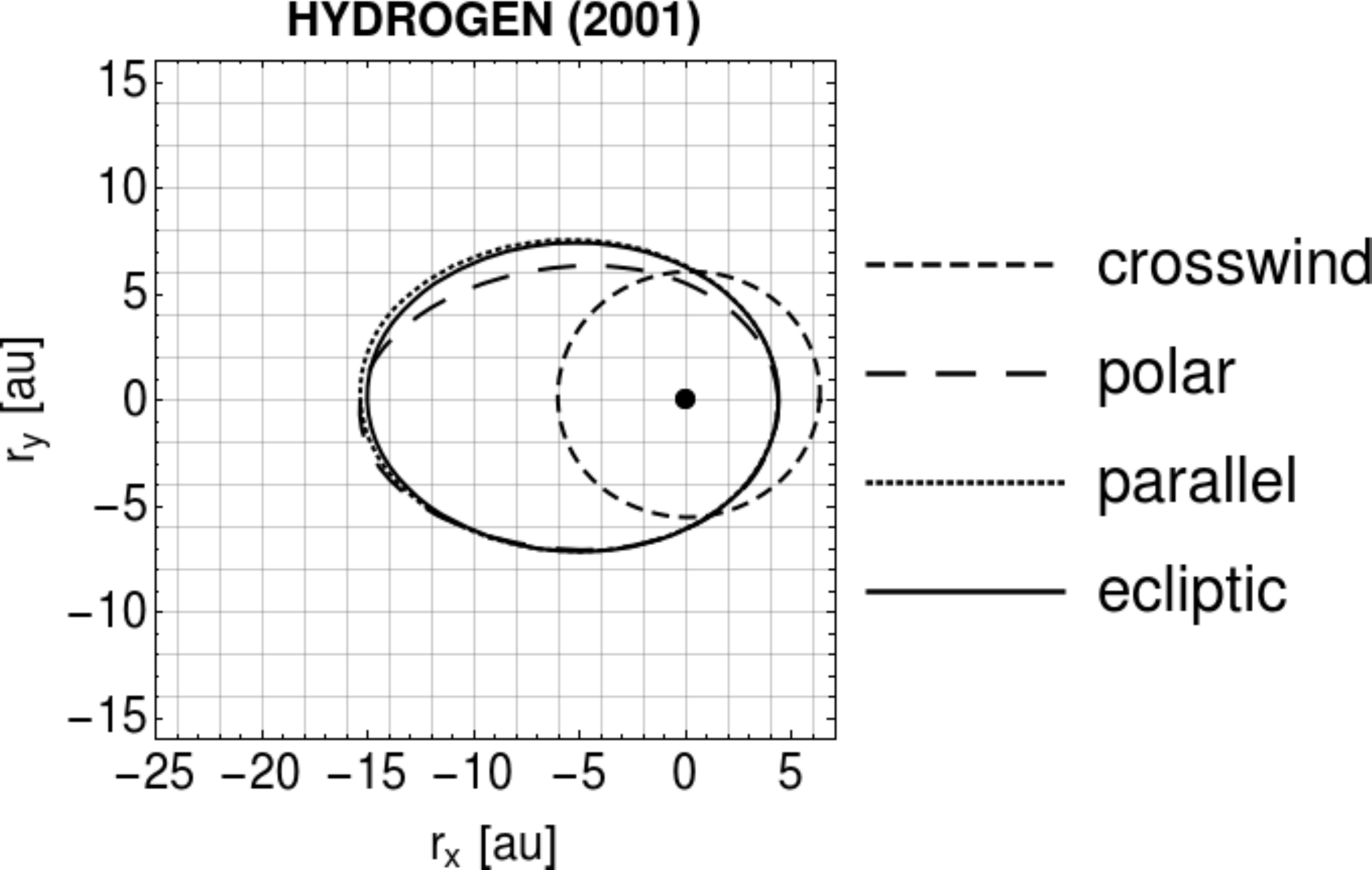} \\
\includegraphics[width=0.3\textwidth]{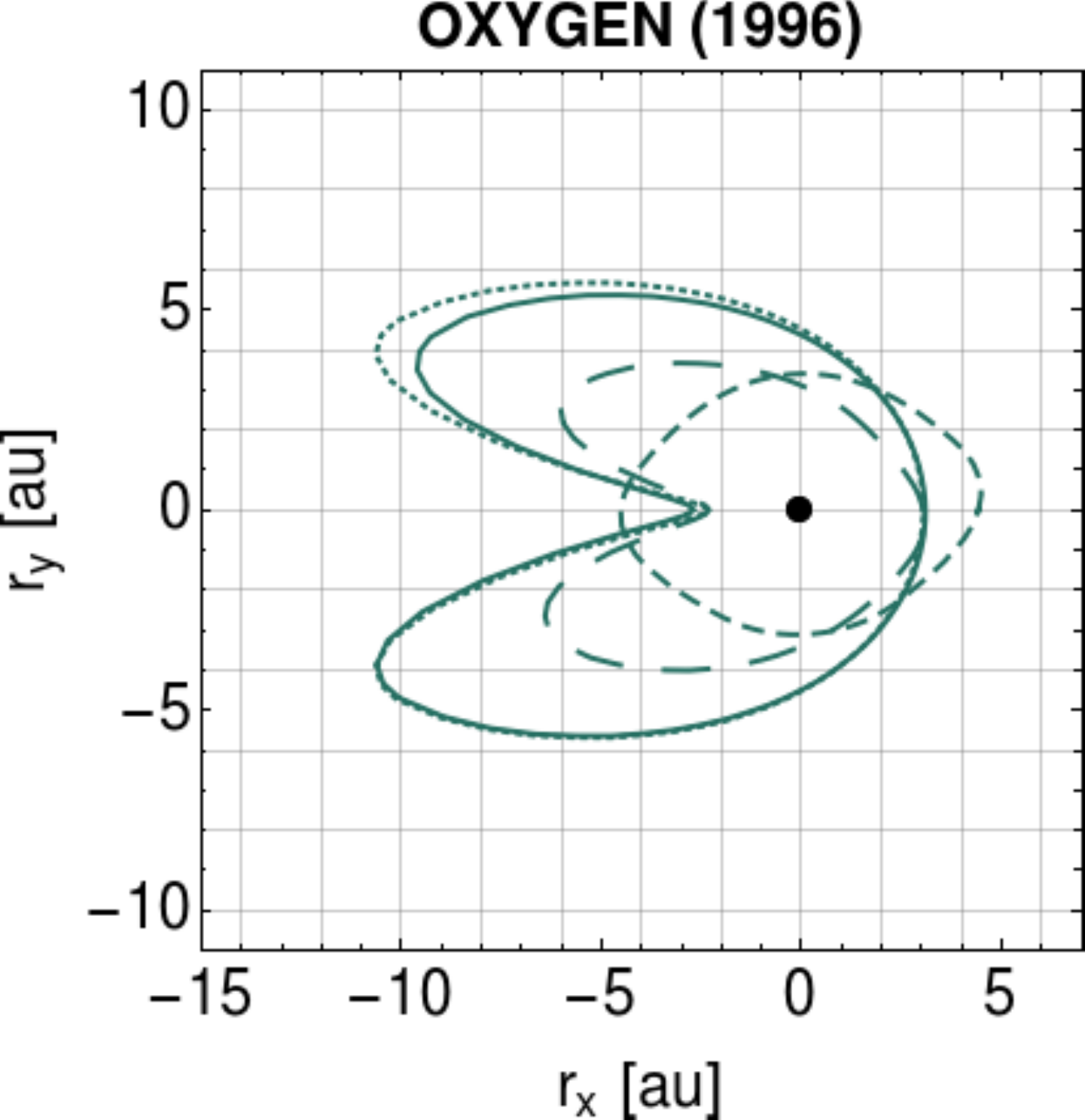} & \includegraphics[width=0.49\textwidth]{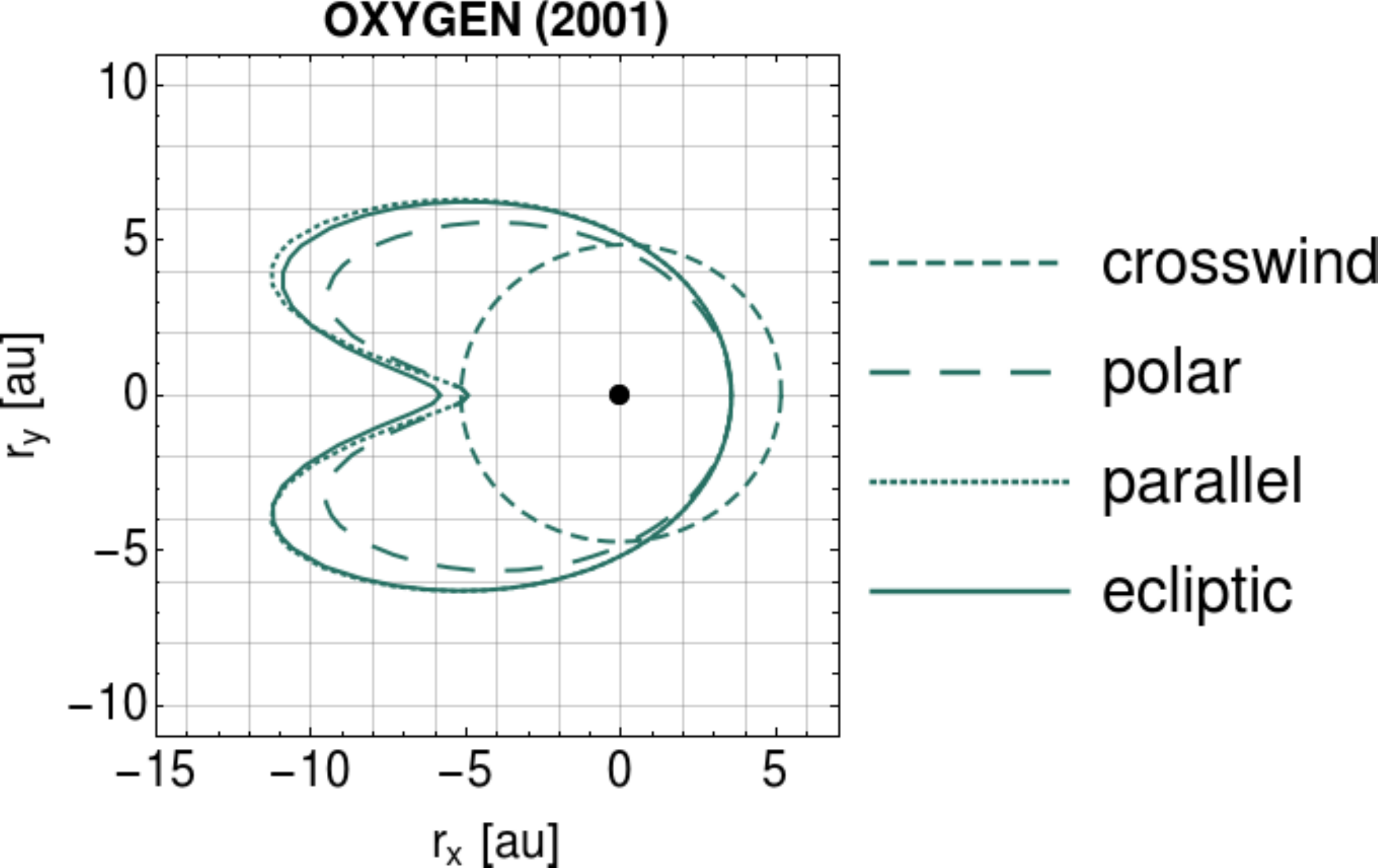} \\
\includegraphics[width=0.3\textwidth]{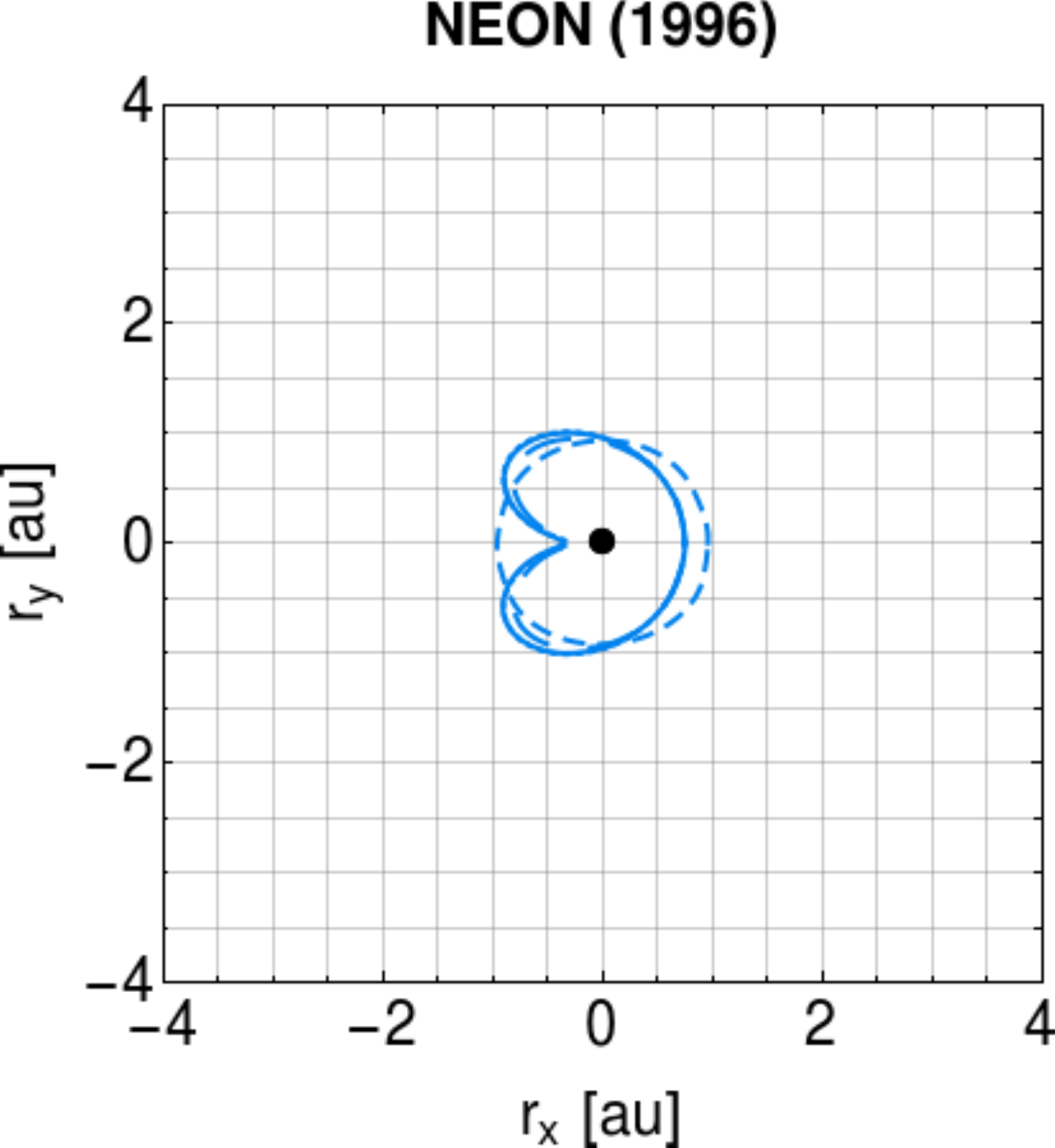} & \includegraphics[width=0.49\textwidth]{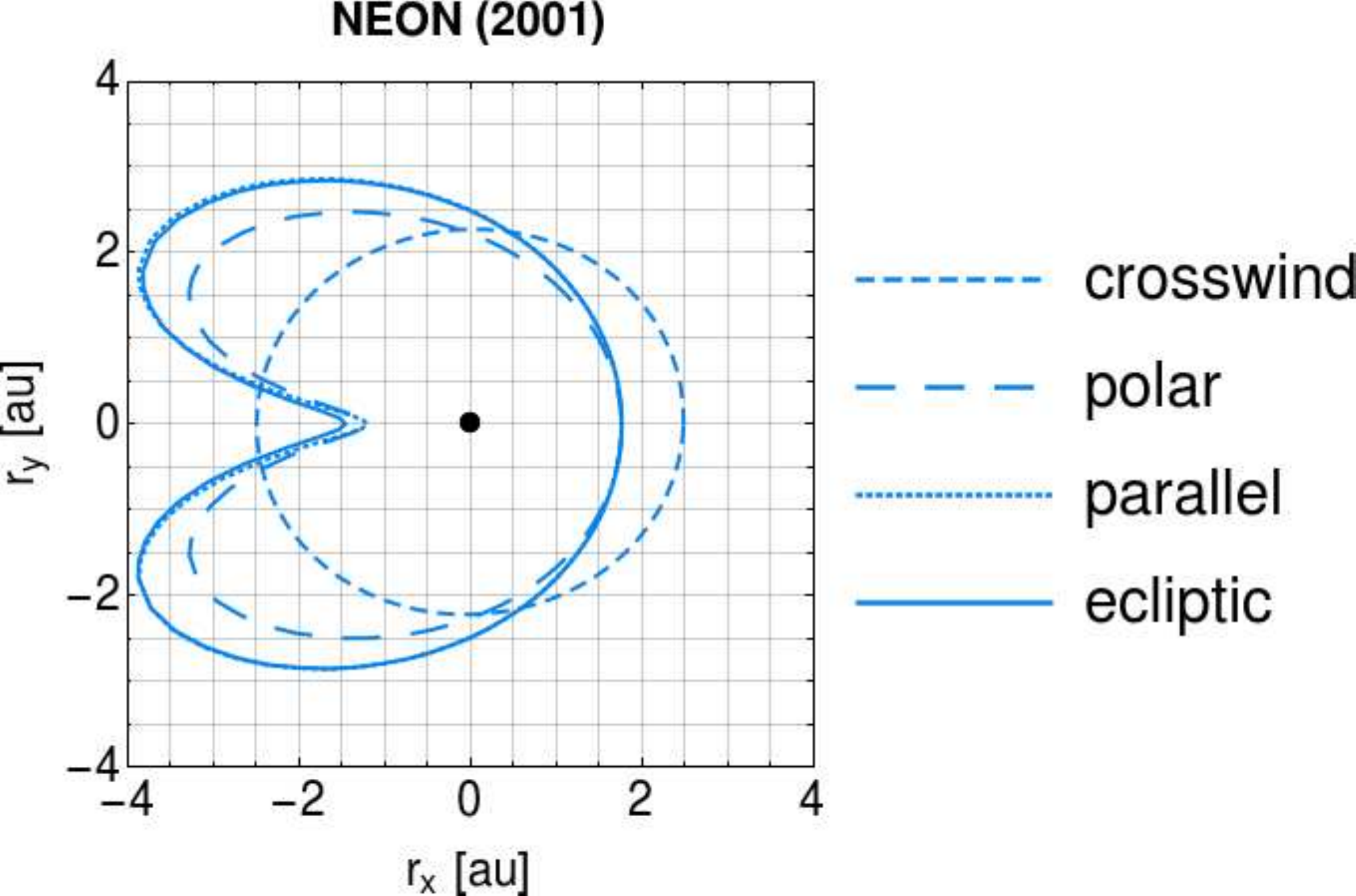} \\
\includegraphics[width=0.3\textwidth]{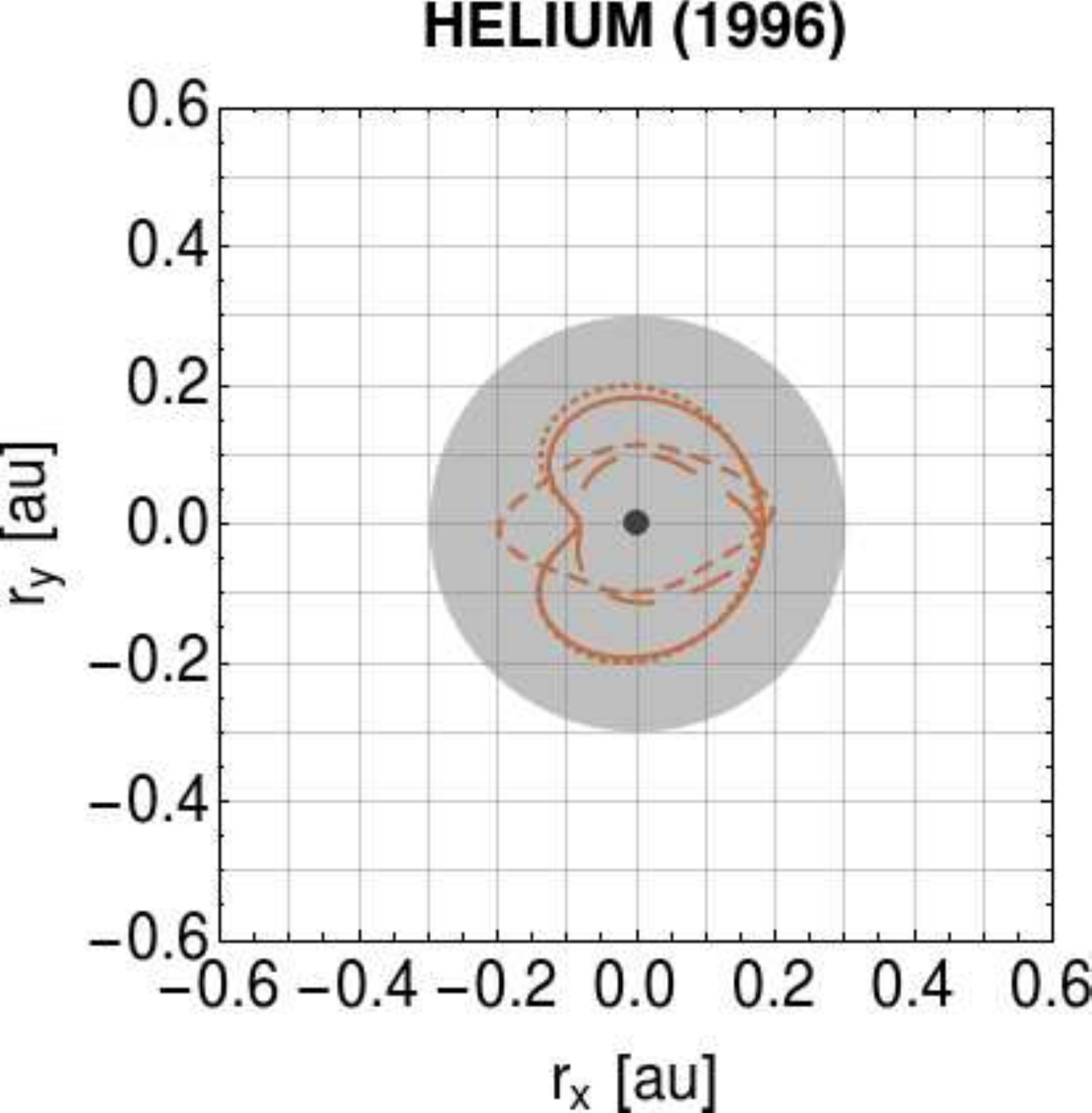} & \includegraphics[width=0.49\textwidth]{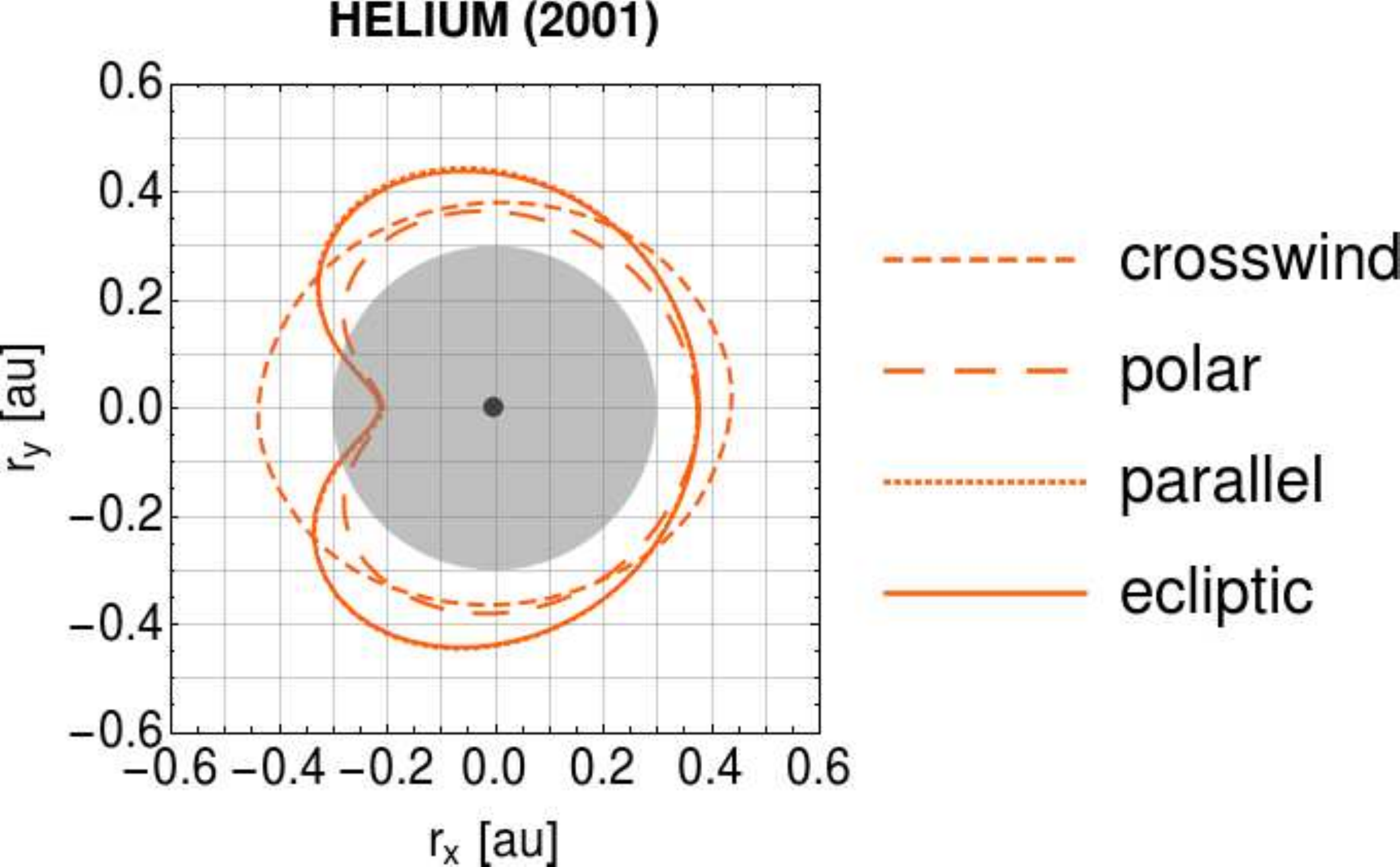} \\
\end{tabular}
\centering
\caption{Cavities of the ISN gas densities for H, O, Ne, and He, defined as a location in space where $n_{\mathrm{ISN,cavity}}=e^{-1}n_{\mathrm{ISN,TS}}$ determined in each of the four planes discussed. Upwind hemisphere is for positive $r_x$ coordinates, and downwind hemisphere for negative $r_x$ coordinates. The left panels show the cavities during the solar minimum in 1996 and the right panels during the solar maximum in 2001. The gray disk in the bottom panels marks the region with distances from the Sun smaller than 0.3~au, within which the ionization rate model used in the calculation is uncertain (see Section~\ref{sec:modelIon}).\label{fig:cavities}}
\end{figure*}

\citet{bzowski_etal:08a} determined the density of ISN~H at TS by analysis of the \textit{Ulysses} H$^+$~PUIs measurements and by taking advantage of the geometrical location of the hydrogen cavity \citep[see also][]{kowalska-leszczynska_etal:18b}. The cavity is by definition a location in space where the local ISN gas density is equal to $e^{-1}$ of the density in the source region, e.g., at TS upwind $\left( n_{\mathrm{ISN,cavity}}=e^{-1} n_{\mathrm{ISN,TS}}\right)$. We use this definition of cavity and determine the locations of the cavities for ISN H, He, O, and Ne. 

Figure~\ref{fig:cavities} shows the determined cavity locations within the crosswind, polar, parallel, and ecliptic planes both during the solar minimum in 1996 and during the solar maximum in 2001. The cavity locations and shapes are significantly different among the species and in various planes. The biggest cavity is for ISN~H and O, the smallest for ISN~He, for which it is located at about 0.5~au in 2001 and closer to the Sun in 1996. The cavity increased for Ne and He in 2001 due to the stronger ionization rates (see the discussion in the previous sections). For ISN~O, the change in cavity location between 1996 and 2001 is small in the parallel and ecliptic planes, in the polar and crosswind planes it is located further away from the Sun in 2001. For ISN~H, the cavity extends to more than 20~au downwind in 1996 and shrinks to about 15~au downwind in 2001. This change for hydrogen cavity between 1996 and 2001 is a result of joint action of both the radiation pressure variations in time and the secular decrease of the solar wind flux between the solar cycles 22 and 23 measured in-situ by instruments in the ecliptic plane and by \textit{Ulysses} out of the ecliptic plane \citep[see, e.g.,][]{mccomas_etal:08a, sokol_etal:13a, sokol_etal:puiIon}. 

Upwind, the ISN~H cavity is located at $\sim4$~au and this location did not change between 1996 and 2001, which is a result of the increase of the solar wind density between 1996 and 2001 as measured by instruments in the Earth's orbit and by \textit{Ulysses}. This is not a typical solar cycle variation, because as discussed by \citet{rucinski_bzowski:95b} (see Figures~8 and 9 there), the variations of the ISN~H cavity location upwind and downwind are about $20\%$ during the solar cycle. For ISN~O, the cavity upwind is located at 3-3.5~au from the Sun. Downwind, however, the ISN~O cavity has a two-wing structure with the closest distance to the Sun along the downwind axis at $\sim2.5$~au and $\sim5$~au in 1996 and 2001, respectively. Further away from the flow axis, the cavity expands up to about 11~au from the Sun. In the case of ISN~Ne, the cavity is located between 0.5 and 1~au with a two-wing structure  similar to that of the ISN~O cavity in the downwind hemisphere in 1996. It expands beyond 1~au and reaches as far as 3~au in the parallel and ecliptic planes for angular distances to the downwind axis of about $45\degr$ in 2001. Upwind, the ISN~Ne cavity moves from $\sim$0.7~au in 1996 to $\sim$1.7~au in 2001. The two-wing shape of the cavity downwind discussed for ISN~O is characteristic also for ISN~Ne, and He. It results from the presence of the focusing cone.

The locations of the cavities for various ISN species together with the successful measurements of the PUIs of this species may allow, in the future, to determine the absolute density of the ISN gas of various species as \citet{bzowski_etal:08a} did for ISN~H.

\subsection{PUI abundances\label{sec:abund}}
\begin{figure*}
\begin{tabular}{cc}
\rotatebox{90}{\textbf{crosswind plane}} & \includegraphics[width=0.95\textwidth]{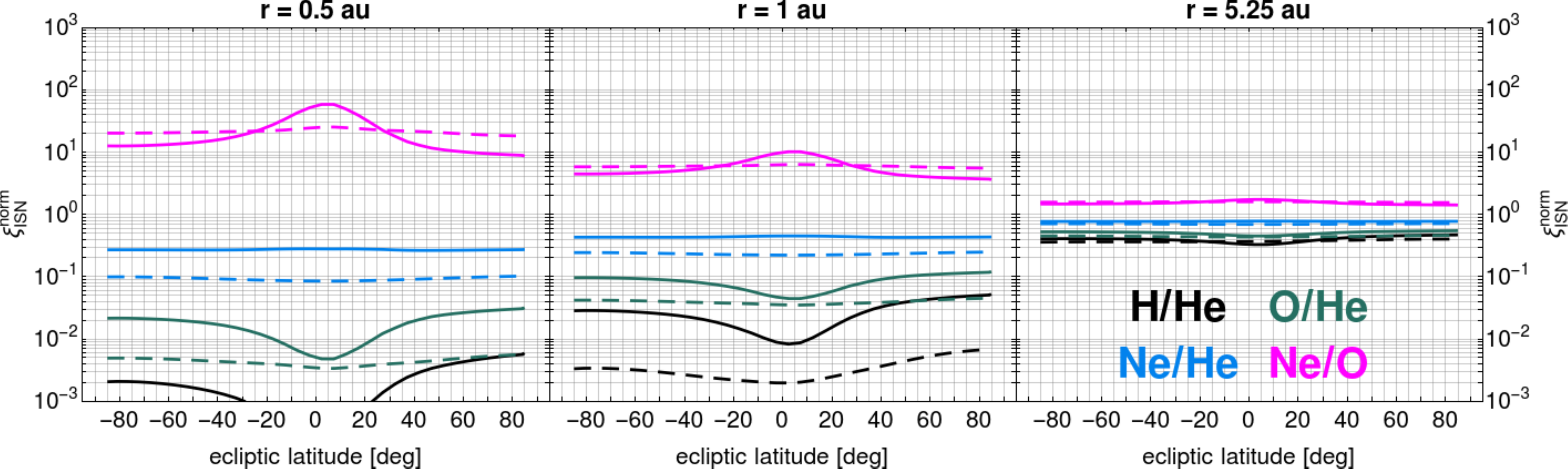} \\
\rotatebox{90}{\textbf{polar plane}} & \includegraphics[width=0.95\textwidth]{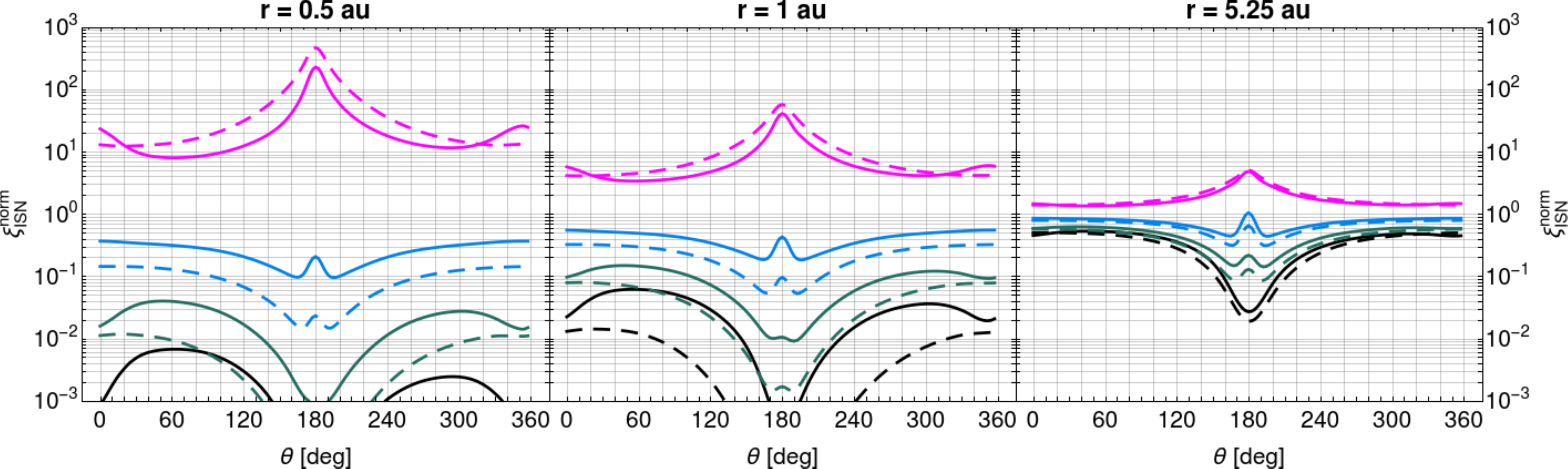} \\
\rotatebox{90}{\textbf{parallel plane}} & \includegraphics[width=0.95\textwidth]{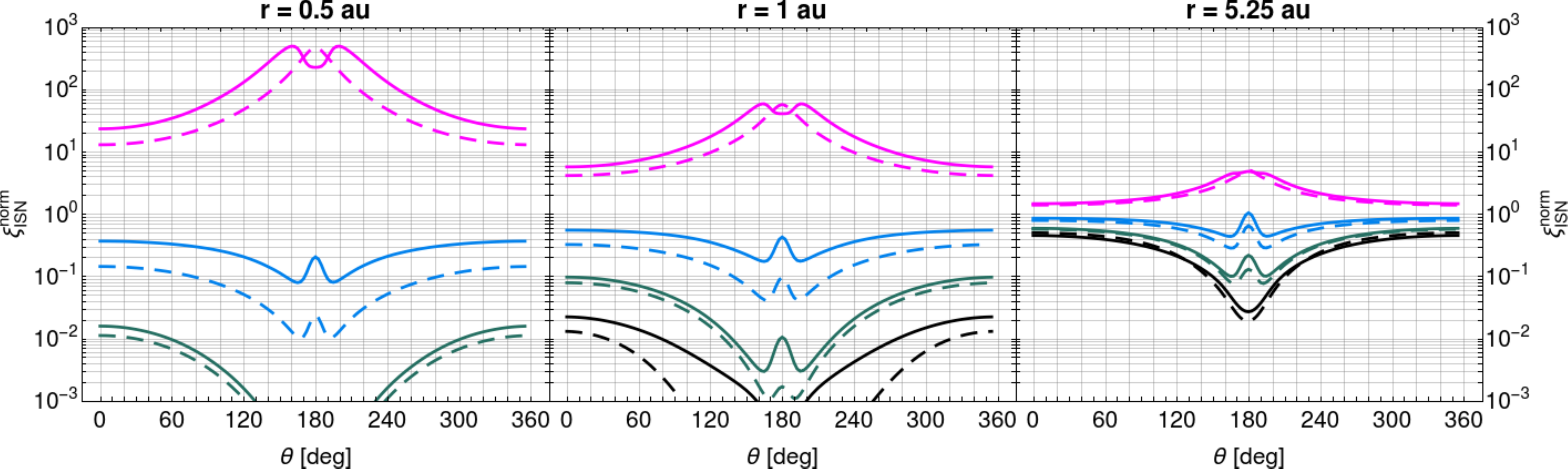} \\
\end{tabular}
\centering
\caption{Relative abundance ratios for ISN gas density calculated according to Equation~\ref{eq:abundISN} along crosswind (top), polar (middle), and parallel (bottom) planes for selected distances from the Sun. The solid lines are for 1996 and the dashed lines are for 2001. \label{fig:abundDensTheta}}
\end{figure*}

\begin{figure*}
\begin{tabular}{cc}
\rotatebox{90}{\textbf{crosswind plane}} & \includegraphics[width=0.95\textwidth]{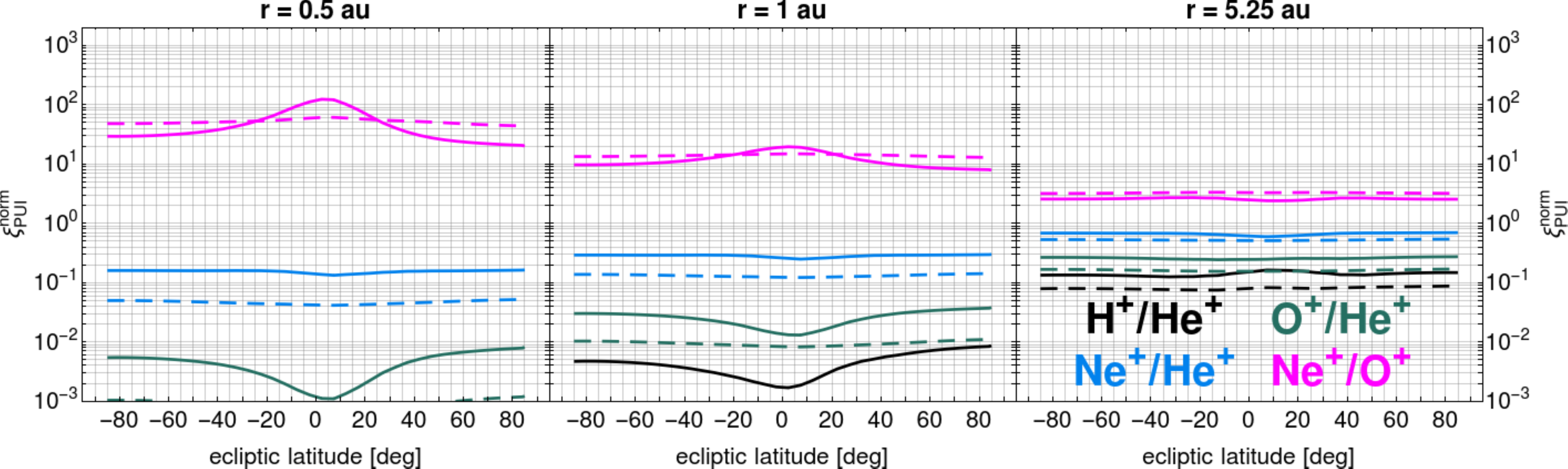} \\
\rotatebox{90}{\textbf{polar plane}} & \includegraphics[width=0.95\textwidth]{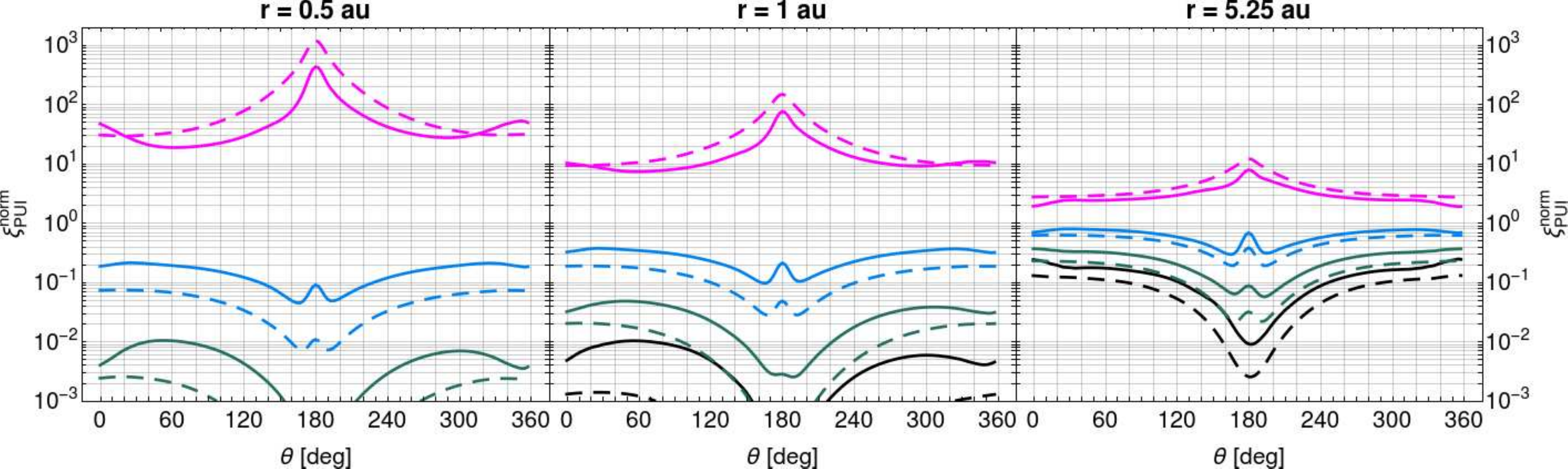} \\
\rotatebox{90}{\textbf{parallel plane}} & \includegraphics[width=0.95\textwidth]{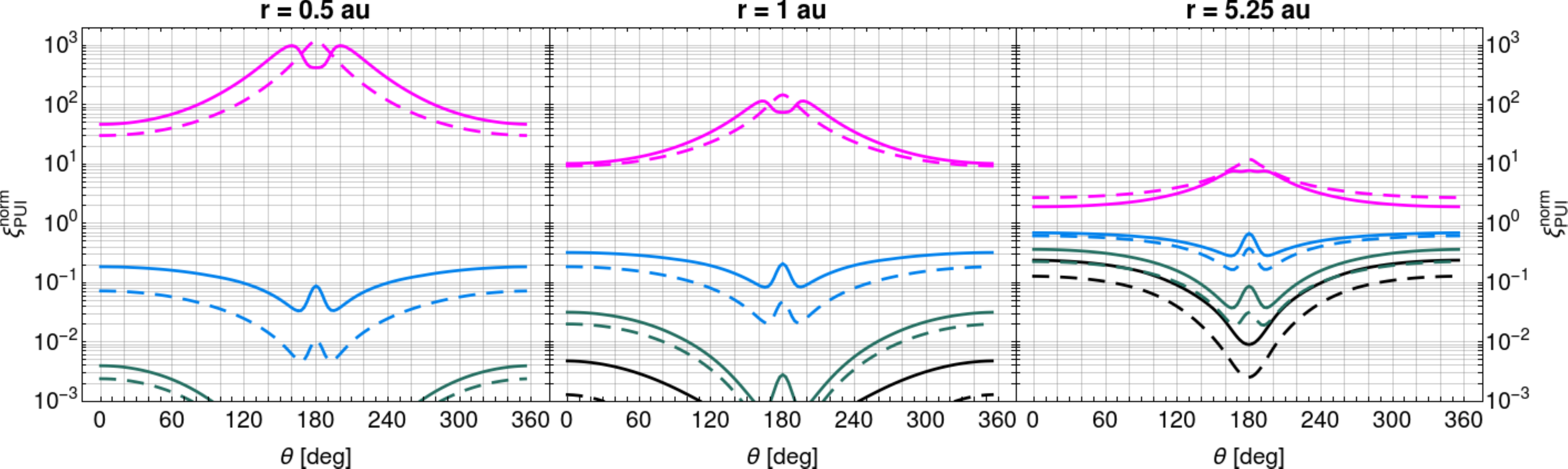} \\
\end{tabular}
\centering
\caption{Same as in Figure~\ref{fig:abundDensTheta}, but for the relative abundance ratios for the PUI density (Equation~\ref{eq:abundPUI}). \label{fig:abundPuiDensTheta}}
\end{figure*}

\begin{table}[t]
\centering
\caption{Variations of the relative abundance ratios of ISN gas ($\xi_{\mathrm{ISN}}^{\mathrm{norm}}$) and PUI ($\xi_{\mathrm{PUI}}^{\mathrm{norm}}$) densities at $r=1$~au with respect to TS ($\xi^{\mathrm{norm}}(r=1)/\xi^{\mathrm{norm}}(r=\mathrm{TS})$) for the upwind (Up) and downwind (Dn) direction for the solar minimum in 1996 and the solar maximum in 2001. \label{tab:abund}}
\begin{tabular}{c|c|c|c|c}
\hline
\multicolumn{5}{c}{$\xi_{\mathrm{ISN}}^{\mathrm{norm}}$} \\
\hline
 	& \multicolumn{2}{c}{1996} & \multicolumn{2}{c}{2001}\\
\hline
 	& Up & Dn & Up & Dn \\
\hline
 Ne/O & $5.6$ & $27.7$ & $4.1$ & $39.0$ \\
 H/He & $2.3 \times 10^{-2}$ & $7.0 \times 10^{-4}$ & $1.3 \times 10^{-2}$ & $6.9 \times 10^{-6}$ \\ 
 Ne/He & $5.5 \times 10^{-1}$ & $2.8 \times 10^{-1}$ & $3.2 \times 10^{-1}$ & $6.3 \times 10^{-2}$ \\
 O/He & $9.9 \times 10^{-2}$ & $1.0 \times 10^{-2}$ & $8.0 \times 10^{-2}$ & $1.6 \times 10^{-3}$ \\
\hline \hline
\multicolumn{5}{c}{$\xi_{\mathrm{PUI}}^{\mathrm{norm}}$} \\
\hline
 & \multicolumn{2}{c}{1996} & \multicolumn{2}{c}{2001} \\
\hline
& Up & Dn & Up & Dn \\
\hline
Ne$^{+}$/O$^{+}$ & $10.2$ & $41.0$ & $5.4$ & $49.2$ \\
H$^{+}$/He$^{+}$ & $4.8 \times 10^{-3}$ & $2.5 \times 10^{-4}$ & $3.0 \times 10^{-3}$ & $2.1 \times 10^{-4}$ \\
Ne$^{+}$/He$^{+}$ & $3.2 \times 10^{-1}$ & $1.5 \times 10^{-1}$ & $1.8 \times 10^{-1}$ & $3.4 \times 10^{-2}$ \\
O$^{+}$/He$^{+}$ & $3.2 \times 10^{-2}$ & $3.7 \times 10^{-3}$ & $3.4 \times 10^{-2}$ & $6.8 \times 10^{-4}$ \\
\hline
\end{tabular}
\end{table}

The composition of the local interstellar matter brings information about the physical state and the chemical evolution of the galactic matter surrounding the Sun. The abundance ratios of the ISN gas species in front of the heliosphere are important for the study of ACRs, for which PUIs are the source population \citep[see, e.g.,][]{chalov_fahr:96a, cummings_stone:96a, gloeckler_etal:09a}. The Ne/O ratio was determined, e.g, by in-situ measurements of the IBEX-Lo detector \citep{bochsler_etal:12a, park_etal:14a}. \citet{bzowski_etal:13b} studied the modulation of the expected abundance ratios of Ne/He, O/He, and Ne/O at 1~au during the solar cycle. Here, we discuss the relative abundance ratios of ISN gas species and PUIs as a function of heliocentric distance and latitude in 1996 and in 2001.

We define the relative abundance ratio for the ISN gas density as: 
\begin{equation}
\xi_{\mathrm{ISN}}^{\mathrm{norm}}=\left( \frac{n_{\mathrm{ISN}}(X)}{n_{\mathrm{ISN}}(Y)} \right)\left( \frac{n_{\mathrm{ISN,TS}}(X)}{n_{\mathrm{ISN,TS}}(Y)} \right)^{-1},
\label{eq:abundISN}
\end{equation}
and for the PUI density as:
\begin{equation}
\xi_{\mathrm{PUI}}^{\mathrm{norm}}=\left( \frac{n_{\mathrm{PUI}}(X)}{n_{\mathrm{PUI}}(Y)} \right)\left( \frac{n_{\mathrm{PUI,TS}}(X)}{n_{\mathrm{PUI,TS}}(Y)} \right)^{-1},
\label{eq:abundPUI}
\end{equation}
where $X$ and $Y$ refer to H, He, Ne, and O, $n_{\mathrm{ISN,TS}}$ and $n_{\mathrm{PUI,TS}}$ are listed in Table~\ref{tab:normF}. We calculate the relative abundance ratios for H/He, Ne/He, O/He, and Ne/O ISN gas densities, as well as for H$^{+}$/He$^{+}$, Ne$^{+}$/He$^{+}$, O$^{+}$/He$^{+}$, and Ne$^{+}$/O$^{+}$ for PUI densities. Effectively, the $\xi^{\mathrm{norm}}$ parameters correspond to the change of a given abundance between the value in the upwind direction at the TS and a given location in space for a given time.

Figure~\ref{fig:abundDensTheta} presents $\xi_{\mathrm{ISN}}^{\mathrm{norm}}$ and Figure~\ref{fig:abundPuiDensTheta} presents $\xi_{\mathrm{PUI}}^{\mathrm{norm}}$ as a function of phase angle $\theta$ along the crosswind, polar, and parallel planes for selected distances from the Sun in 1996 and in 2001. As presented, the modulation of the relative abundance ratios is from one to three orders of magnitude between TS and 1~au.

The Ne/O and Ne$^{+}$/O$^{+}$ abundance ratios increase with respect to the TS with the decrease of the heliocentric distance. The respective ratios to He decrease toward the Sun. This is a result of stronger depletion of the ISN~Ne and O density than the ISN~He density inside the heliosphere (Figure~\ref{fig:densTheta}) due to higher ionization rates for Ne and O than for He (see Figures~3 and 8 in \citet{sokol_etal:puiIon}). The relative abundance ratios for ISN gas density and PUI density vary with the angle with respect to the ISN gas flow direction, as illustrated in the parallel and polar planes in Figures~\ref{fig:abundDensTheta} and \ref{fig:abundPuiDensTheta}. The highest ratios are expected downwind for Ne/O and Ne/He (and also for Ne$^{+}$/O$^{+}$ and Ne$^{+}$/He$^{+}$), for the ratios of O/He (O$^{+}$/He$^{+}$) and H/He (H$^{+}$/He$^{+}$), the highest magnitudes are expected upwind. Table~\ref{tab:abund} summarizes the variation of $\xi_{\mathrm{ISN}}^{\mathrm{norm}}$ and $\xi_{\mathrm{PUI}}^{\mathrm{norm}}$ between 1~au and TS for the upwind and downwind directions in 1996 and 2001. 

The double-peak structure for Ne/O and Ne$^+$/O$^+$ in 1996 and the double-minimum structure for Ne/He (Ne$^+$/He$^+$) and O/He (O$^+$/He$^+$) in both 1996 and 2001 at small angular distances to the downwind axis, present in Figure~\ref{fig:abundDensTheta}, result from different modulation of the ISN gas and PUI densities as a function of phase angle at close angular distances to the downwind axis. As illustrated already in Figures~\ref{fig:densTheta} and \ref{fig:densPUITheta}, the ISN~He and He$^+$~PUI density smoothly increase from upwind to downwind, whereas the density for ISN~Ne and O and Ne$^+$ and O$^+$~PUIs decrease from upwind to downwind with a minimum at about $20\degr$ from the downwind axis and next significantly increase as a cone. This results in double-peak/minimum variations of the relative abundance ratios for the ISN gas and PUIs as a function of angular distance from the downwind direction. Such structures might have important implications for interpretation of measurements, as the measured ratio may decrease while approaching to the downwind axis with a subsequent rapid increase.

\subsection{Anistropy of ISN gas distribution \label{sec:anisotropy}}
\begin{figure*}[t!]
\begin{tabular}{cc}
\includegraphics[width=0.41\textwidth]{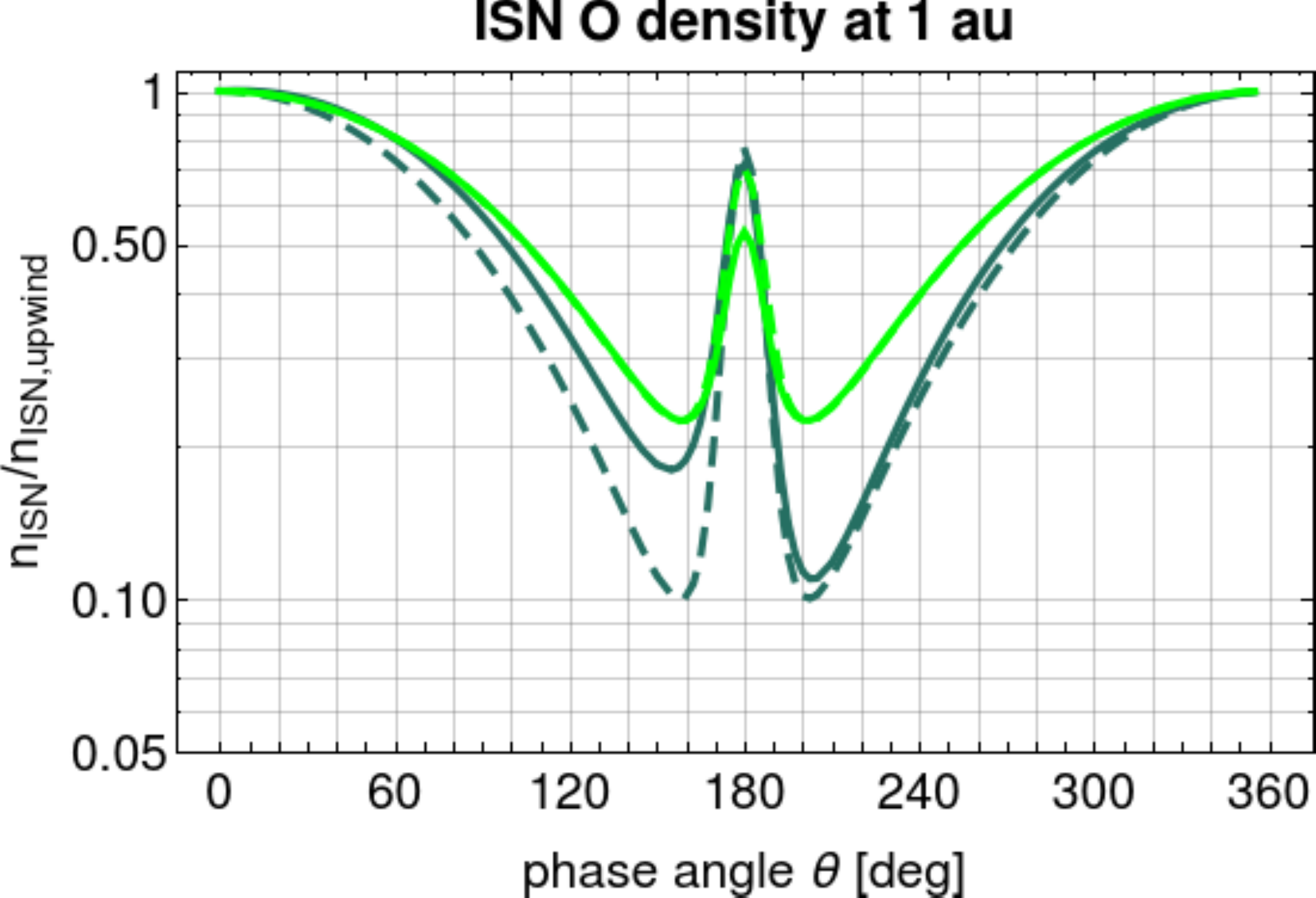} & \includegraphics[width=0.6\textwidth]{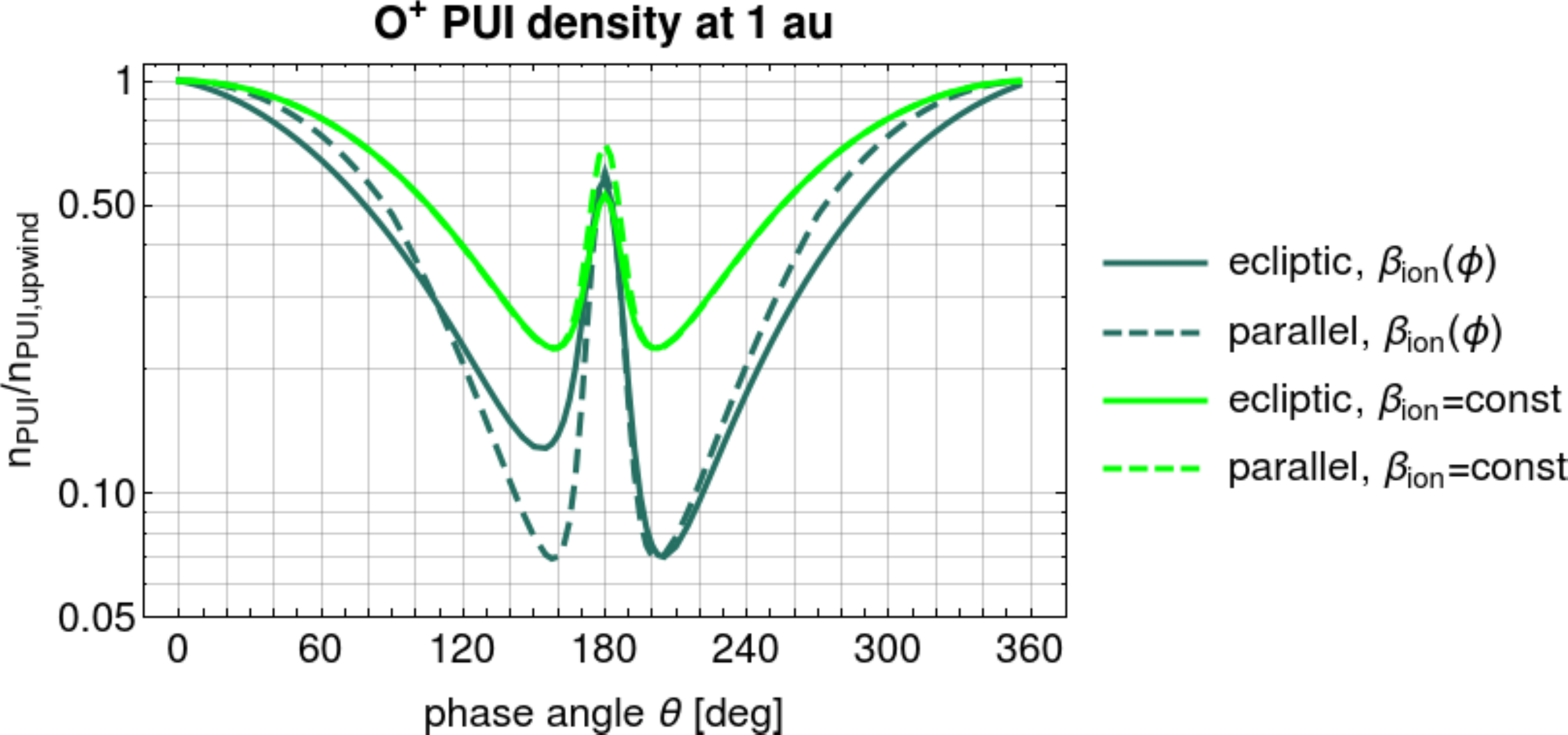} \\
\end{tabular}
\centering
\caption{Anisotropy of the ISN O gas density (left) and O$^{+}$~PUI density (right) in the ecliptic (solid lines) and parallel (dashed) planes at 1~au in 1996. The dark green lines illustrate the case with ionization rates variable in heliographic latitude, while the light green lines illustrate a case with the ionization rates constant in time and latitude equal to $5\times10^{-7}$~s$^{-1}$ (see the text for an explanation). Please note that the light green solid and dashed lines are almost identical except in the downwind direction. \label{fig:anisoO}}
\end{figure*}

The ISN atoms traverse various latitudes before being detected in the ecliptic plane, especially for the downwind hemisphere. The latitudinal variation of the ionization rates results in significant departures of the ISN gas density from axial symmetry along the upwind-downwind direction, and consequently of the PUI density. We compare the ISN gas and PUI densities expected at 1~au for both ecliptic and parallel planes for the case of oxygen, calculated with the total ionization rates constant in time and spherically symmetric ($\beta_{\mathrm{ion}}=5\times10^{-7}$~s$^{-1}$, which is a total ionization rate value characteristic for the solar minimum for ISN~O in the ecliptic plane and an average value for the polar latitudes, see Figure~8 in \citet{sokol_etal:puiIon}) with those for the ionization rates variable in time and in heliographic latitude (see results in Figure~\ref{fig:anisoO}). 

Oxygen is the most prone for the anisotropies of the solar ionizing medium via the solar wind, which contributes to the charge exchange and electron impact ionization reactions (as discussed by \citet{sokol_etal:puiIon}). For the case of constant ionization rates, in both ecliptic and parallel planes the ISN~O gas density variations as a function of phase angle are almost identical in shape (an exception is the downwind direction), with about $80\%$ reduction of the gas at an angular distance to the cone of $\sim20\degr$ with respect to the upwind direction. A narrow cone is formed inside, where the gas density increases to about $0.5$ and $0.7$ of the value upwind in the ecliptic and parallel planes, respectively. However, the ISN~O densities calculated for the more realistic case, with the ionization rates anisotropic in heliographic latitude, show asymmetries in the downwind hemisphere. The asymmetries are well visible in the ecliptic plane, which is not in phase with the variation of the heliographic latitude along the plane as it is in the parallel plane (see bottom panel in Figure~\ref{fig:planePhAng}). While the magnitudes of the ISN density in the cone are similar in both planes ($\sim0.72$ and $\sim 0.77$ of the value upwind in the ecliptic and the parallel planes, respectively), the reduction of the gas density before the cone is greater for the parallel plane, and after the cone both in the parallel and ecliptic planes, with a reduction as much as $90\%$.

These anisotropies in the ISN gas propagate to the PUI density, being further modulated by the asymmetric ionization rates that produce ions. The right-hand panel in Figure~\ref{fig:anisoO} presents the O$^+$~PUI densities calculated for the same assumptions regarding ionization rates and ISN densities as in the left-hand panel. Again, for the case with the ionization rates constant in time and latitude, the PUI densities between the ecliptic and parallel planes differ only in the cone, with smaller reduction with respect to the upwind direction in the parallel plane (about $30\%$ reduction) than in the ecliptic plane (about $50\%$ reduction). In the case with ionization rates variable in time and heliographic latitude, the anisotropy as a function of phase angle is present along the ecliptic plane. The amplitude between the smallest PUI density before the cone and after the cone to the value in the cone varies from $\sim0.22$ to $\sim0.12$, respectively. 

The anisotropy of the ISN gas and PUI densities along the ecliptic plane is present for all species; however, the effect is the strongest for oxygen, and next for neon, for which the gas and PUI densities in the cone are about $30\%$ higher in the parallel plane. Thus, the anisotropy of the PUI variations along the ecliptic plane (if measurable), together with careful modeling of the ISN gas modulation inside the heliosphere, might serve as a complementary tool to study the latitudinal variations of the ionization rates that modulate both ISN gas and PUIs, at least for ISN~O.

\subsection{Variation at TS \label{sec:varTS}}
\begin{figure*}
\begin{tabular}{cc}
\includegraphics[width=0.4\textwidth]{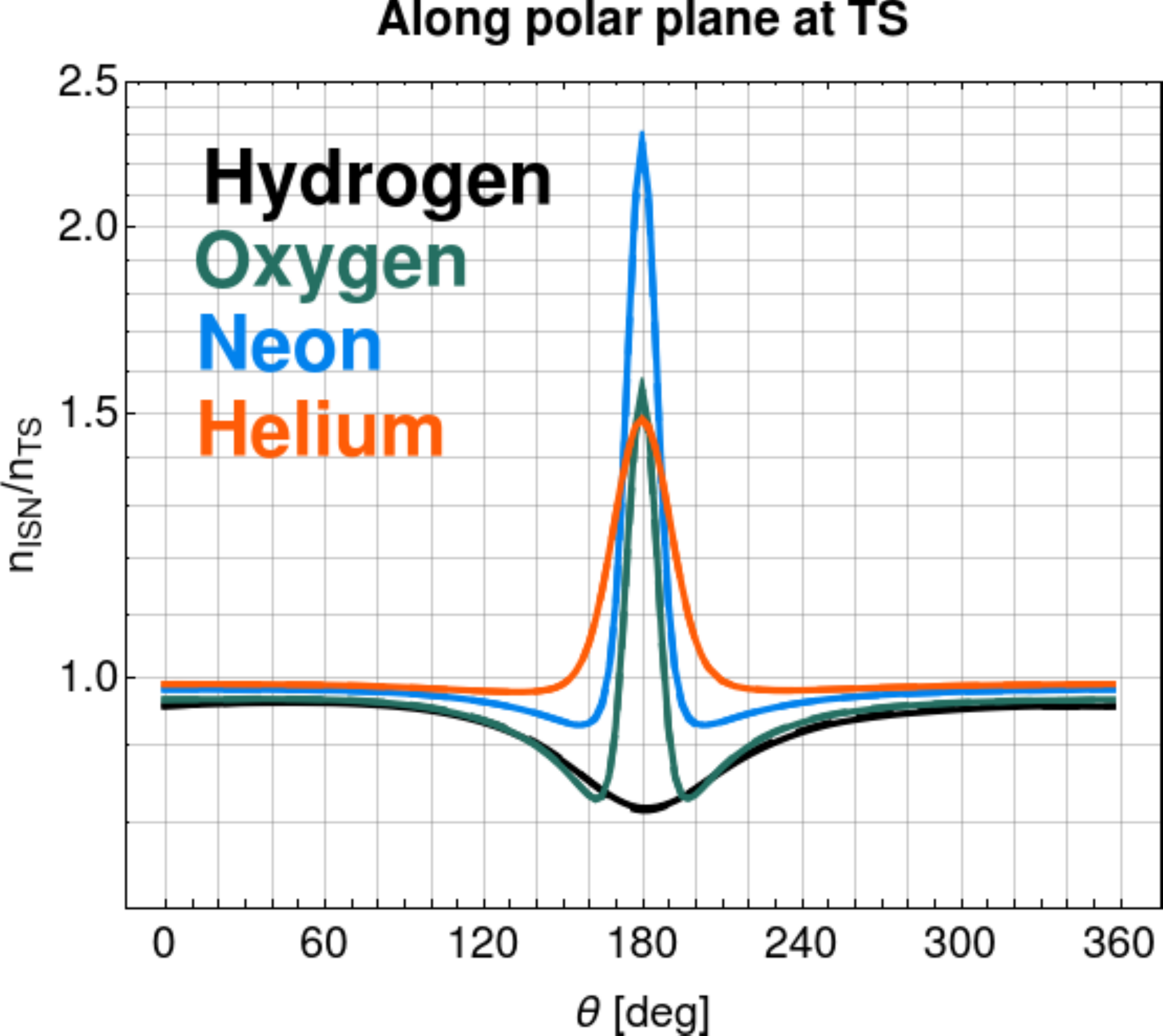}  & \includegraphics[width=0.41\textwidth]{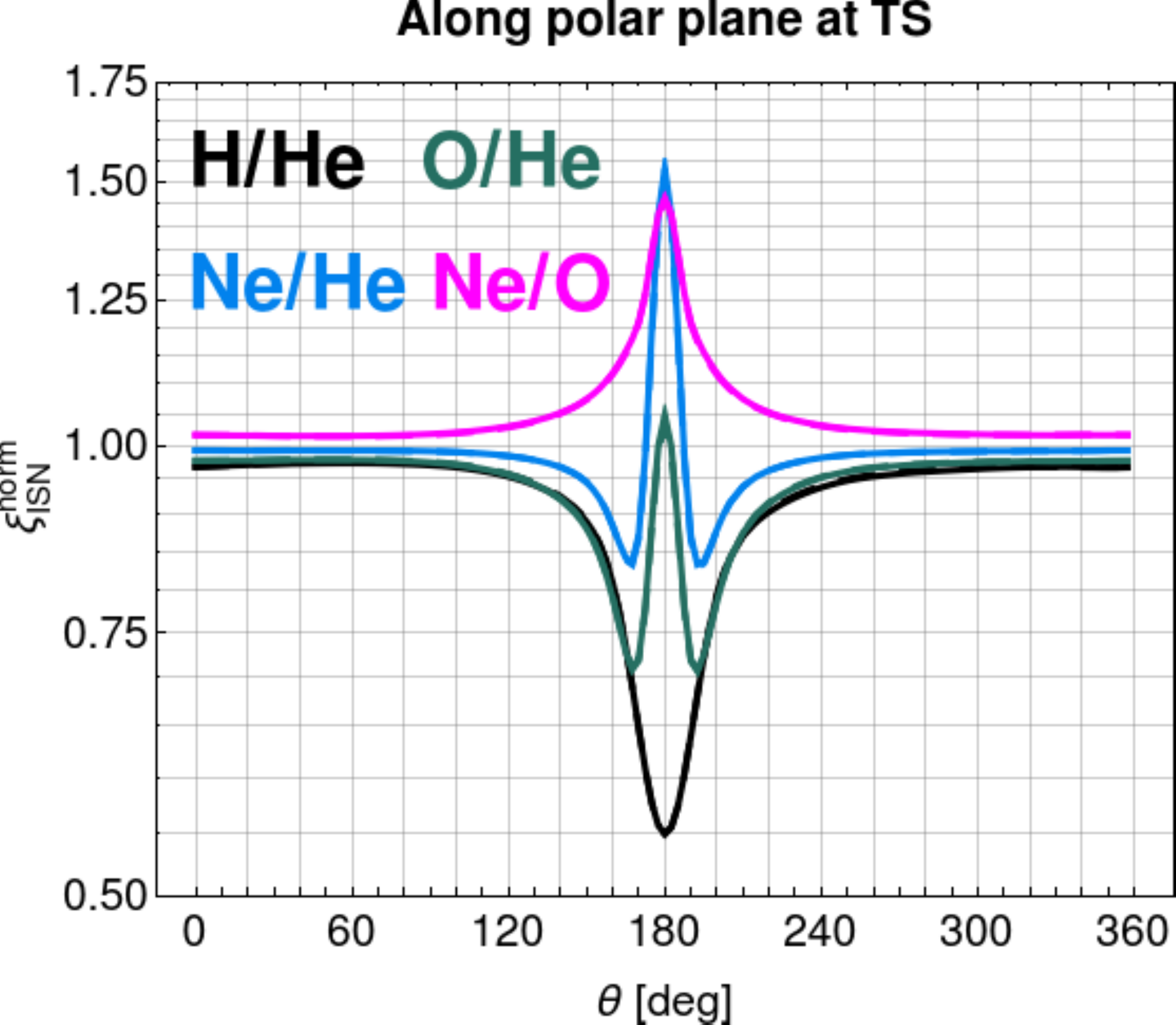}  \\
\end{tabular}
\centering
\caption{Variation of the normalized ISN gas density (left) and the ISN gas density relative abundance ratios (right) as a function of phase angle $\theta$ along the polar plane at TS. \label{fig:atTS-isn}}
\end{figure*}

\begin{figure*}
\begin{tabular}{c}
\includegraphics[width=0.95\textwidth]{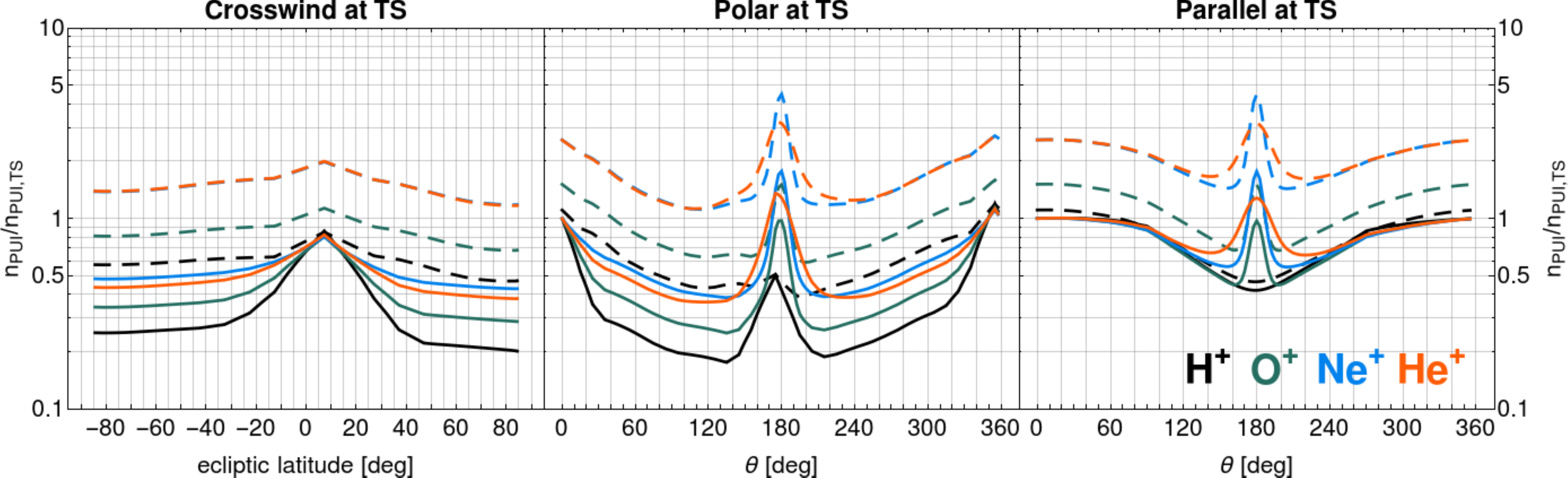} \\ \includegraphics[width=0.95\textwidth]{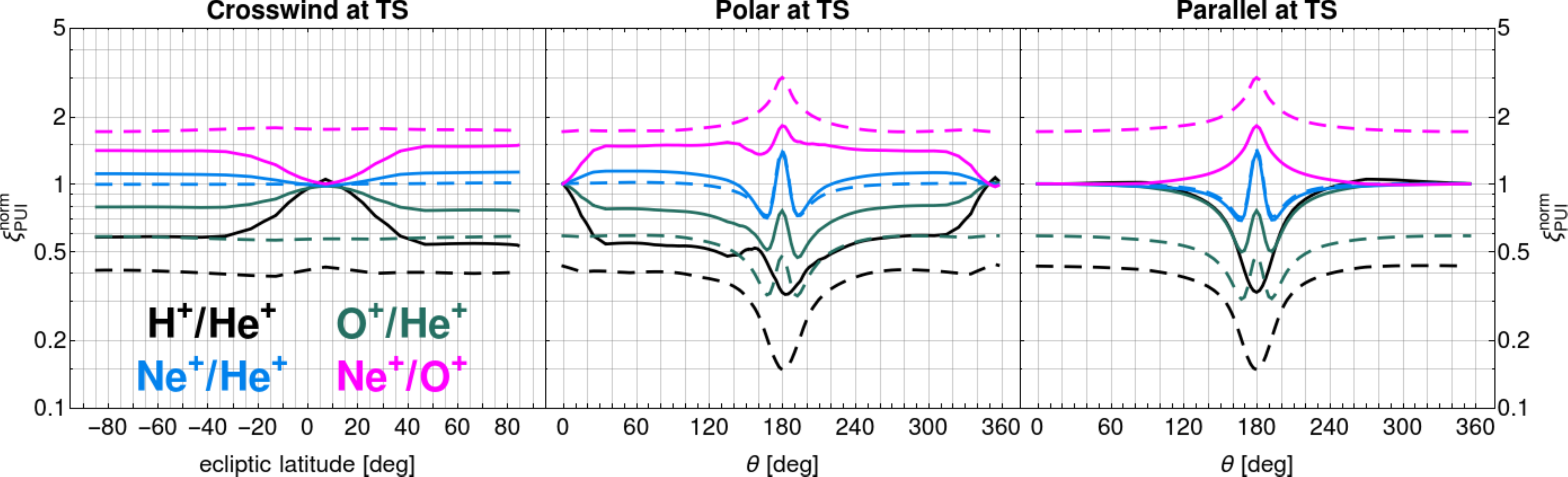} \\
\end{tabular}
\centering
\caption{Variation of the normalized PUI density (top) and the PUI density relative abundance ratios (bottom) along the crosswind, polar, and parallel planes at the TS in 1996 (solid lines) and in 2001 (dashed lines). \label{fig:atTS-puis}}
\end{figure*}

The cones of ISN gas and PUI densities that form in the downwind hemisphere persist up to the TS and beyond, as presented in Figure~\ref{fig:atTS-isn}. The ISN gas density remains constant along the TS up to about $40\degr$ from and after the downwind axis, where the change of the density with respect to upwind is greater than $10\%$. The strongest modification from the upwind value is exactly downwind, where the ISN~gas density is $\sim2.3$ for Ne and $\sim1.5$ for He and O greater than upwind. For H, the density downwind is about 0.8 of the density upwind. Consequently, the relative ISN gas density abundance ratios (Equation~\ref{eq:abundISN}) also diverge from the upwind value, as illustrated in Figure~\ref{fig:atTS-isn}. The H/He ratio decreases downwind to $\sim0.55$ of the upwind value. The Ne/O and Ne/He ratios are 1.5 higher than upwind, and the smallest increase downwind is for the O/He ratio being just about 1.05; however, the increase follows a decrease by about $30\%$ at $\sim15\degr$ from the downwind axis. A similar decrease just before the increase is present for the Ne/He ratio, being about $20\%$. No significant differences in the ISN gas variations along the TS are expected between 1996 and 2001.

The modulation of PUI densities at the TS is much more intense than the modulation of the ISN gas. Figure~\ref{fig:atTS-puis} presents the variation of the normalized PUI density and the relative PUI density abundance ratios along the crosswind, polar, and parallel planes in 1996 and in 2001. The PUI densities vary significantly as a function of phase angle from the upwind direction. For H$^+$~PUI, a decrease downwind is formed, being about $50\%$ of the upwind value. For the PUIs of the remaining species, the highest PUI densities are again expected downwind at TS; the increase is more than $30\%$ for He$^+$ and about $70\%$ for Ne$^+$. 

In contrast to the ISN gas densities, the PUI densities at the TS strongly vary between 1996 and 2001. The PUI densities are greater during the solar maximum than during the solar minimum for all species and in all three planes considered. For H$^+$~PUIs, the greatest increase is for the polar plane, since the whole latitudinal structure changes due to the change of the latitudinal variations of the ionization rates with the solar activity cycle. The greatest increase of the PUI density at the TS due to the change of the phase of the solar activity is for He$^+$ and Ne$^+$, being about a factor of 2.5 for upwind to 3 (He$^+$) and 4 (Ne$^+$) downwind. For O$^+$~PUIs, the density at the TS in 2001 is about 1.5 times greater than in 1996 both upwind and downwind.

The nonuniform spatial distribution and solar cycle variations of the PUIs at the TS are reflected in the PUI density relative abundance ratios (Equation~\ref{eq:abundPUI}), as illustrated in the bottom row panels in Figure~\ref{fig:atTS-puis}. The $\xi_{\mathrm{PUI}}^{\mathrm{norm}}$ remains close to 1 only in the parallel plane and to about $40\degr$ angular distance from the downwind axis on both sides. In the parallel plane, the Ne$^+$/O$^+$ ratio increases downwind by a factor of 1.8 and the H$^+$/He$^+$ ratio decreases by a factor of almost 0.3. In the case of Ne$^+$/He$^+$ and O$^+$/He$^+$ ratios, the modulation with phase angle in the parallel plane initially shows a decrease (up to 0.7 and 0.5 of the upwind value, respectively), and next an increase downwind (to $\sim0.7$ and $\sim1.4$, respectively). This spatial anisotropy of the PUIs might be reflected in the ACRs if only observations for various directions and moments in time were available.

\subsection{Preferable locations for observations \label{sec:maxLoc}}
\begin{figure*}
\includegraphics[width=\textwidth]{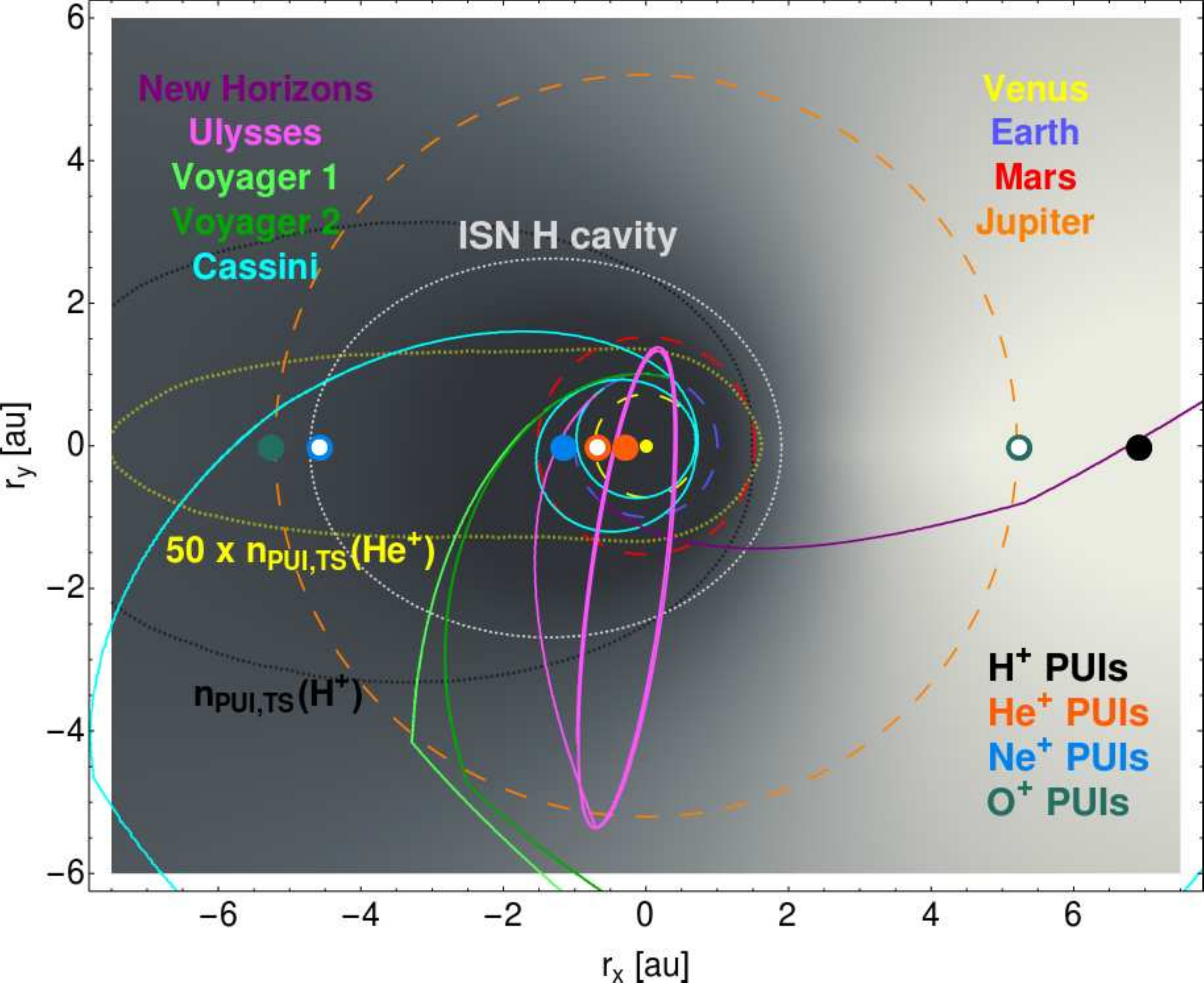}
\centering
\caption{Locations of the maxima of PUI density of H$^{+}$ (black), He$^{+}$ (orange), Ne$^{+}$ (blue), and O$^{+}$ (green) in 1996 (filled dots) and in 2001 (empty dots) in the ecliptic plane. The locations of the maxima for the H$^{+}$~PUI in 1996 and 2001 are the same within the grid adopted in calculations. The H$^{+}$ PUI density in 1996 is presented as a background (see Figure~\ref{fig:contoursPUIDensMinLin}) with the upwind/downwind hemisphere for the positive/negative $r_x$ coordinates of distance from the Sun, respectively. The orbits of Earth, Venus, Mars, and Jupiter and the trajectories of \textit{Ulysses}, \textit{Voyager}~1 and \textit{Voyager}~2, \textit{New Horizons}, and \textit{Cassini} are superimposed. The ISN~H cavity in 1996 is marked by the light gray dotted line. The yellow dotted line encircles a region inside which the He$^{+}$~PUI density is 50 times greater than upwind at the TS in 1996. The black dotted line encircles a region outside which the H$^{+}$~PUI density is greater than the density upwind at the TS in 1996. The yellow dot in the middle of the graphics is the Sun (not to scale). \label{fig:masterPUI}}
\end{figure*}

PUIs originating from the ISN gas have been important tools to diagnose the processes inside and outside of the heliosphere since their direct detection by SULEICA/AMPTE \citep{mobius_etal:85a}. Since then, PUIs have been successfully measured by many missions, e.g., \textit{Ulysses} \citep{gloeckler_etal:93a, gloeckler_geiss:98a, cannon_etal:14a}, \textit{Cassini} \citep{mccomas_etal:04a, hill_etal:09a}, \textit{New Horizons} \citep{mall_etal:98a, randol_etal:13a, mccomas_etal:17b}, \textit{Voyager} 1 \& 2 \citep{argall_etal:17a, hollick_etal:18a, hollick_etal:18b}, ACE \citep{gloeckler_etal:98a, gloeckler_geiss:01a, chen_etal:13a, fisher_etal:16a}, \textit{STEREO} \citep{drews_etal:10a, drews_etal:12a, drews_etal:15a, taut_etal:18a}, \textit{MESSENGER} \citep{gershman_etal:14a}, \textit{SOHO}/CELIAS \citep{berger_etal:15a, taut_etal:15a}. 

Figures~\ref{fig:contoursPUIDensMinLogr} and \ref{fig:PUIDensInflowRadial} present that detection at 1~au is preferable for He$^+$~PUIs regardless of the phase of the solar activity, which has been confirmed by observations (e.g., ACE, \citet{fisher_etal:16a}). Also the maximum of the Ne$^+$~PUI density is expected around 1~au during the solar minimum. However, for H$^+$ and O$^+$~PUIs, as well as for Ne$^+$~PUIs during the solar maximum, the preferable locations for detection, defined by the maximum of the PUI density, are at distances from about 5~au up to about 25~au. Thus, as this study shows, it is challenging to sample PUIs of various species with one instrument at certain locations, because of the spatial separation of the PUI density maxima (Figures~\ref{fig:masterPUI}). 

As already pointed out by \citet{hollick_etal:18b}, and illustrated in Figures~2 and 3 in their paper, \textit{Voyager} 1 \& 2 spacecraft trajectories after the launch transitioned from a region of dominance of the He$^+$~PUIs production rate to a region where the H$^+$~PUI production rate dominates in the heliosphere. The transition where more H$^+$ than He$^+$ PUIs are produced, happens between $\sim3 - 4$~au. If only the fluxes of Ne$^+$ and O$^+$ PUIs are high enough, the Ne$^+$ PUIs should be observed by \textit{Voyager} just after launch, and O$^+$ PUIs around Jupiter orbit. However, Table~\ref{tab:normF} shows that the expected densities for Ne$^+$ and O$^+$ are from 2 to 3 orders of magnitude lower than the densities of He$^+$ and H$^+$.

For H$^+$~PUIs, the most interesting location is the upwind hemisphere, where the ISN~H gas is not strongly depleted by the ionizing solar radiation (Figure~\ref{fig:masterPUI}). Advantage of this was taken by the \textit{New Horizons} observations \citep{mccomas_etal:17b}. The downwind hemisphere is the most preferable region to sample PUIs of He$^+$, Ne$^+$, and O$^+$, see Figure~\ref{fig:masterPUI} and also \citet{mall_etal:96a, mall_etal:96b, mall_etal:98a}. The studies by \textit{Ulysses} \citep{gloeckler:96a}, \textit{Cassini} \citep{mccomas_etal:04a, hill_etal:09a}, \textit{Voyager} 1\& 2 \citep{hollick_etal:18a, hollick_etal:18b, hollick_etal:18c}, and \textit{New Horizons} \citep{mccomas_etal:17b} measurements confirmed the detection of H$^+$ and He$^+$~PUIs along the spacecraft trajectories, while \citet{geiss_etal:94a} reported also about the detection of Ne$^+$ and O$^+$~PUIs by \textit{Ulysses}. Interestingly, as our study shows, for the O$^+$~PUIs both the upwind and downwind hemispheres seem to be good locations for observations; however, the most preferable distances from the Sun are from Jupiter up to Pluto orbits.

\section{Summary \label{sec:Summary}} 
We studied the ISN gas density for H, He, Ne, and O and the H$^+$, He$^+$, Ne$^+$, and O$^+$ PUI density spatial distributions from inside 1~au up to the TS during the solar minimum and maximum. Our study shows that different species have different ISN and PUI density structures because of different modulation by the solar ionizing factors. We showed that the latitudinal anisotropy of the ionization rates causes anisotropy of the ISN gas density and PUI density measured along the ecliptic plane. Because the anisotropy is different for various species, the ISN and PUI density relative abundance ratios vary nonuniformly in space and in time.

The study showed that while ISN density maxima are expected outside 1~au, the PUI density maxima are expected closer to the Sun. The greatest PUI densities through the heliosphere are expected for He$^+$~PUIs, next for H$^+$~PUIs, and next for Ne$^+$ and O$^+$ PUIs. We concluded that simultaneous observations of PUIs of all of the discussed species might be challenging because of the spatial separation of the PUI density maxima inside the heliosphere. The most preferable location for the detection of He$^+$~PUIs is downwind inside 1~au. For Ne$^+$~PUIs the best moment for detection is the solar minimum at 1~au, while during the solar maximum the peak is shifted almost to the Jupiter's orbit with more than $50\%$ reduction of the density. The O$^+$~PUIs could be looked for both upwind and downwind, for which the intensities are expected to be similar; however, the acceptable locations are at distances starting from the Jupiter orbit up to a few tens of astronomical units. The upwind hemisphere is confirmed to be the best location to detect H$^+$~PUIs.

\acknowledgments
The authors thank Jacob Heerikhuisen for sharing the model of the heliosphere that allowed them to assess the distance to the heliospheric termination shock, and Charles Smith for encouraging them to publish the first ideas of the study developed here. J.M.S. thanks Marcin Szpanko for the help with preparation of Figure~\ref{fig:masterPUI} and Duncan Pettengill (Wolfram Technology Group) for technical support with Wolfram Research Mathematica software. The presented study is supported by the Polish National Science Center grant No. 2015/19/B/ST9/01328.

\begin{figure*}
\includegraphics[width=\textwidth]{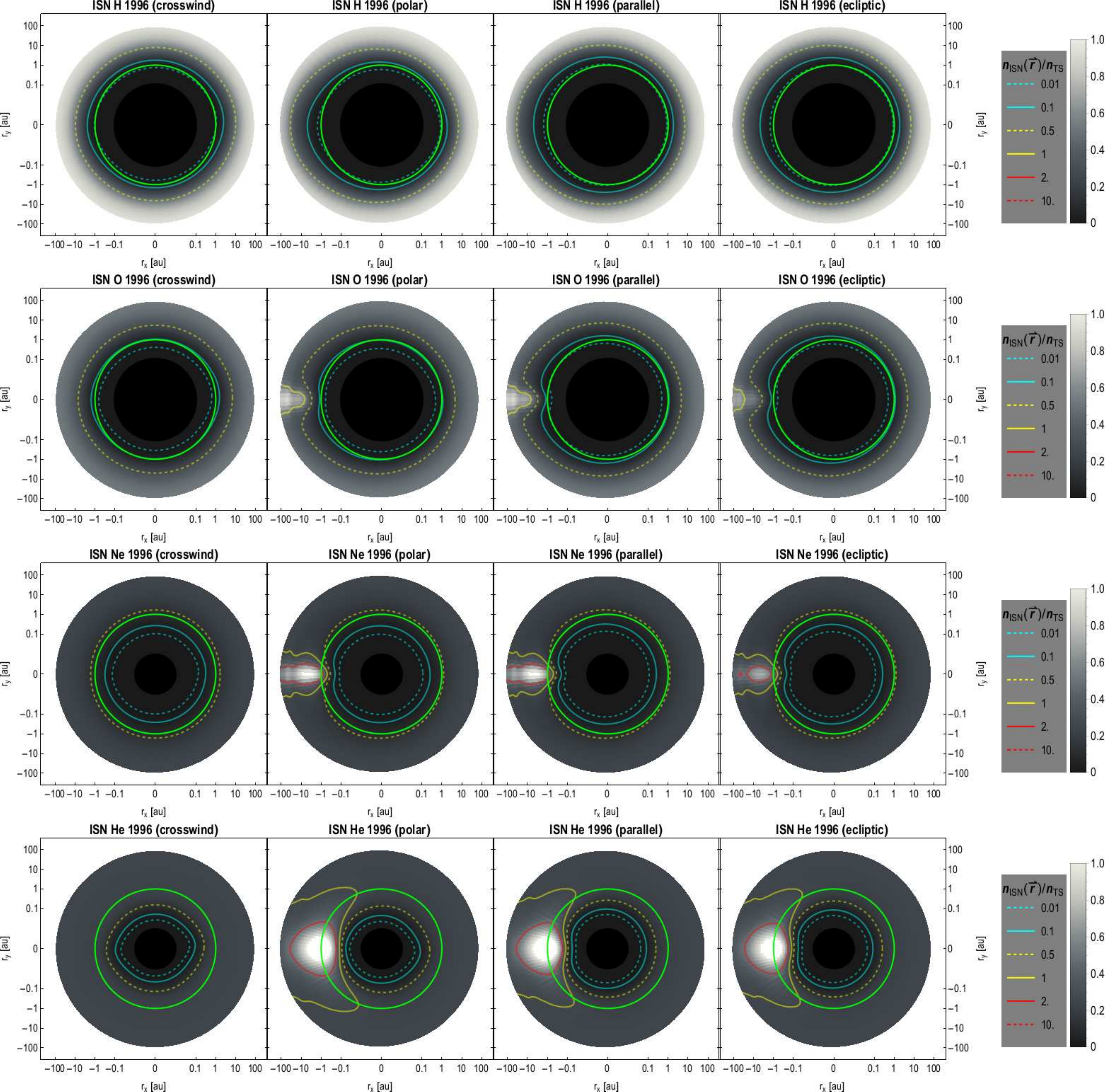}
\centering
\caption{Same as in Figure~\ref{fig:contoursDensMinLin} but with logarithmic scale in the distance to the Sun. The distances smaller than 0.3~au are not studied because of not enough information about the ionization rates inside 0.3~au (see text) and thus masked by the black disks. \label{fig:contoursDensMinLogr}}
\end{figure*}

\bibliographystyle{apj}
\bibliography{sokol_PUIsGlobal_accepted_accepted_2arXiv}{}

\end{document}